\documentclass{iopart}
\usepackage{soul}		

\usepackage{iopams}

\usepackage{graphicx}
\usepackage{epstopdf}
\usepackage{xcolor}
\usepackage{sidecap}
\usepackage{booktabs}		
\usepackage{colortbl}		
\usepackage{multirow}		
\usepackage{multicol}		
\usepackage[sort&compress,comma,numbers]{natbib}

\newcommand{\rone}{\ensuremath{r_{\rm I}}}
\newcommand{\rtwo}{\ensuremath{r_{\rm II}}}
\newcommand{\gangle}{\ensuremath{\varphi}}
\newcommand{\kB}{\ensuremath{k_{\rm B}}}		

\newcommand{\Dyzero}{\ensuremath{\Delta y_0}}		
\newcommand{\sx}{\sigma_x}				
\newcommand{\sy}{\sigma_y}				

\begin{document}

\title[]{Controlling strongly correlated dust clusters \\with lasers}

\author{Hauke Thomsen$^1$, Patrick Ludwig$^1$, Michael Bonitz$^1$, Jan Schablinski$^2$, Dietmar Block$^2$, Andr\'e Schella$^3$ and Andr\'e Melzer$^3$}

\address{$^1$ Institut f\"ur Theoretische Physik und Astrophysik, Christian-Albrechts-Universit\"at zu Kiel, 24098~Kiel, Germany}
\address{$^2$ Institut f\"ur Experimentelle und Angewandte Physik, Chrsitian-Albrechts-Universit\"at zu Kiel, 24098~Kiel, Germany}
\address{$^3$ Institut f\"ur Physik, Ernst-Moritz-Arndt-Universit\"at Greifswald, 17487~Greifswald, Germany }
\ead{\mailto{thomsen@theo-physik.uni-kiel.de}, \mailto{ludwig@theo-physik.uni-kiel.de}, \mailto{bonitz@physik.uni-kiel.de}, \mailto{schablinski@physik.uni-kiel.de}, \mailto{block@physik.uni-kiel.de},
\mailto{schella@physik.uni-greifswald.de}, \mailto{melzer@physik.uni-greifswald.de}}
\begin{abstract}
Lasers have been used extensively to manipulate matter in a controlled way -- from single atoms and molecules up to macroscopic materials. They are particularly 
valuable for the analysis and control of mesoscopic systems such as few-particle clusters. Here we report on recent work on finite size complex (dusty) plasma systems.
These are unusual types of clusters with very strong inter-particle interaction so that, at room temperature, they are practically in their ground state. 
Lasers are employed as a tool to achieve excited states and phase transitions.

The most attractive feature of dusty plasmas is that they allow for a precise diagnostic with single-particle resolution.
From such measurements, the structural properties of finite
two-dimensional (2D) clusters and three-dimensional (3D) spherical crystals in nearly harmonic
traps--so-called Yukawa balls--have been explored in great detail. Their structural features--the shell
compositions and the order within the shells--have been investigated and good agreement to
theoretical predictions was found. Open questions on the agenda are the excitation behavior,
the structural changes, and phase transitions that occur at elevated temperature.

Here we report on recent experimental results where laser heating methods were further improved and applied to finite 2D and 3D dust clusters.
Comparing to simulations, we demonstrate that laser heating indeed allows to increase the temperature in a controlled manner.
For the analysis of thermodynamic properties and phase transitions in these finite systems, we present theoretical and experimental results on the basis of the instantaneous normal modes, pair distribution function and the recently introduced center-two-particle correlation function.

\end{abstract}

\maketitle

\section{Introduction}\label{s:intro}

Complex (dusty) plasmas differ from traditional plasmas in a number of aspects: 
complex plasmas, in very general terms, are multicomponent plasmas that contain, 
in
addition to electrons, ions and neutral atoms, also (macro-)molecules.

This may lead to substantial chemical reactivity or growth of clusters or nanoparticles
that 
are of high interest for technological applications, see e.g.~\cite{loewen_book12,meichsner_recent_2012,bonitz_introduction_2010,bonitz_complex2_2014}. 
Alternatively,
macroscopically large particles (dust) can be introduced 
into the plasma externally which
may radically alter the plasma properties.
These ``dusty plasmas'' have evolved into a separate research field and are the subject of this review. 
Here, we focus on non-reactive dusty plasmas containing comparatively large---typically micrometer sized---monodisperse plastic spheres.

The unusual properties of these plasmas arise from the behavior of the dust particles and their enormously high charge. In the plasma, the particles are subject to continuous bombardment by the much lighter electrons, ions and neutrals. 
In a radio frequency (rf) discharge, the
electron temperature is way above the ion temperature resulting in a higher impact rate of
electrons onto the originally neutral particles, compared to the ions. As a consequence the
particles become highly negatively charged, with the charge reaching values on the order of $Q_{\rm d}
= \mathcal{O}(-10\,000\,e)$ elementary charges for a particle of micron size~\cite{melzer_experimental_1994}.
It is due to this high negative charge that the dust particles are strongly interacting with each other, and the dust component of the plasma aquires a very large electrostatic energy that may exceed the thermal energy by orders of magnitude:
the dust becomes {\em strongly coupled} (or strongly correlated), in striking contrast to usual high-temperature plasmas.
At the same time electrons and ions are only weakly coupled and can often be regarded as a more or less uniform background. 

Such strong correlation effects are presently of high interest in a large variety
of fields, including condensed matter, dense plasmas (such as warm dense matter), ultracold
quantum gases or the quark-gluon plasma. In fact, dusty plasmas serve as a prototype for studying
correlation phenomena, e.g. \cite{bonitz_complex_2010}. A particular advantage of dusty plasmas is the 
large particle size and large inter-particle spacing (on the order of several micrometers) which allows for a direct optical imaging of individual particles. At the same time the large mass results in slow characteristic time scales (on the order of hundreds of milliseconds), so the whole dynamics of these plasmas can be studied on the single-particle ``atomic'' level~\cite{loewen_book12,bonitz_introduction_2010}. Despite the purely classical behavior, dusty plasmas allow for unique and unprecedentedly accurate diagnostics of strong correlation effects which are important for other strongly correlated systems where such diagnostics are missing.

To have a quantitative measure of correlation effects in these highly non-ideal charged particle systems, it has
been common to use the Coulomb coupling parameter $\Gamma$ that relates the mean
Coulomb interaction energy of two particles to the thermal energy as\footnote{Strictly, the definition (\ref{eq:Coulomb_Gamma}) is an appropriate measure for the potential energy only when the pair interaction is Coulombic. In the case of screening, the effective
coupling parameter depends on the screening length $\lambda_{\rm
D}$~\cite{ott_non-invasive_2011}. For the results reported in this paper, the dust-dust
interaction is moderately screened and the definition (\ref{eq:Coulomb_Gamma}) is sufficiently
accurate.
}
\begin{equation}
  \Gamma = \frac{Q_{\rm d}^2}{4\pi\varepsilon_0\ b_{\rm WS}} \frac{1}{\kB T_{\rm d}}\ {\rm ,}
 \label{eq:Coulomb_Gamma}
\end{equation}
where the Coulomb interaction is estimated by the one of two particles separated by the
Wigner-Seitz radius, that is related to the density by $n^{-1}=4\pi b^3_{\rm WS}/3$ in 3D
($n^{-1}=\pi b_{\rm WS}^{2}$ in 2D\footnote{For macroscopic systems, the Coulomb coupling parameter is clearly defined by the condition
that each particle occupies, on average a spherical volume (a circle in 2D) with the radius
equal to $b_{\rm WS}$.}), and $T_{\rm d}$ is the kinetic temperature that corresponds
to the dust particles' random motion. In ideal (or weakly non-ideal), conventional plasmas $\Gamma \ll
1$. If $\Gamma$ exceeds unity, the system is strongly coupled and the particle arrangement
exhibits an increasingly long range order, giving rise to liquid-like behavior. If $\Gamma$ exceeds 
about one hundred the interaction is strong enough to spatially localize the particles leading to 
a crystalline structure. In macroscopic Coulomb systems, the critical value for the freezing transition 
is around $\Gamma=175$ in 3D and $\Gamma=137$ in 2D~\cite{ichimaru_strongly_1982, PhysRevE.72.026409, bonitz_classical_2008}.

Extended (nearly macroscopic) dust systems in 2D and 3D have been realized in experiments for decades, see e.g. Refs.~\cite{Chu94b,thomas_plasma_1994,Hayashi94,Morfill99b,Rothermel02,Schmidt11}. 
There, $\Gamma$ values of several hundreds are easily achieved, even at room temperature, as will be discussed in Secs.~\ref{sec:2Dexp} and \ref{s:3d}.
At these conditions the system is practically in its ground state.
Therefore---and this may be surprising---for dusty plasmas, the major challenge for many applications is to {\em lower the coupling strength}.
The main task is to do this in a controlled manner for instance by increasing only the dust temperature $T_{\rm d}$, without affecting the other plasma parameters.
This is particularly important in order to study the thermodynamic properties, the melting process and the different phases of these systems.
To achieve a controlled heating of dust clusters several methods have been used, including variation of the rf power or the neutral gas pressure~\cite{melzer_experimental_1996,Thomas96,ivlev_melting_2003,Couedel10}.
However, these methods usually alter the whole discharge environment and, thus, effectively create a different plasma, making a ``clean'' analysis of dust thermal effects difficult~\cite{ivanov_melting_2005,melzer_experimental_1996}.

\subsection{Laser heating of dust in the context of laser-matter interaction}\label{ss:laser-matter}
A suitable approach to heat the dust without affecting the discharge environment is the use of
lasers which is, therefore, in the focus of the present review. Since laser manipulation has become 
common in many fields it is of interest to put the present activities in the field of dusty plasmas into a broader perspective,
before moving on.
 
The control of matter by lasers has seen dramatic progress over the last two decades which is due to the rapid increase of 
available coherent radiation sources. These are, in first place, optical and infrared lasers but, with the progress in the field of harmonic generation 
as well as of free electron lasers, also high photon energies---from UV to soft x-rays---became available.
At the same time a tremendous variety of methods and mechanisms to control matter with lasers is being utilized which is briefly summarized 
in table \ref{tab:laser-matter}. There we list the main properties of coherent electromagnetic radiation and how they are applied in different areas.
\begin{table}
 \begin{tabular}{|l||l|}
    \hline
    {\bf Laser property}                 & {\bf Applications and Mechanisms}	  \\ \hline\hline
    {\bf photon energy}	                 & $\bullet$ (multi-)photon excitation and ionization of atoms, \\
    $N\hbar \omega, \quad N=1, 2, \dots$ & $\quad$ molecules, condensed matter  etc. \\ 
                                         & $\bullet$ photoemission spectroscopy \\ 
                                         & $\bullet$ inverse bremsstrahlung heating of matter \\
                                         & $\bullet$ {\bf photoionization of dust particles} \\\hline
    laser polarization                   & $\bullet$ selection of intra-atomic (intra-molecular) transitions, \\
                                         & $\quad$ e.g. linear vs. circular polarization \\\hline
    electrical field strength  $E$,      & $\bullet$ tunnel and field ionization of atoms \\               
    potential energy $-eEx$              & $\bullet$ field assisted creation of electron-positron pairs \\
                                         & $\quad$ (Schwinger mechanism) \\\hline
    laser intensity, $I \sim E^2$        & $\bullet$ heating, melting, evaporation  of matter            \\
    Ohmic heat                           & $\bullet$ ionization of matter, creation of dense plasmas        \\\hline
    {\bf light pressure}, $p\sim I$      & $\bullet$ structuring of surfaces, micro-hole boring\\
                                         & $\bullet$ {\bf acceleration of single dust particles}\\\hline
    short pulse duration                 & $\bullet$ time-resolved diagnostics (e.g. pump-probe schemes)\\
                                         & $\bullet$ rapid excitation of non-equilibrium phenomena\\
                                         & $\bullet$ non-thermal melting of solids \\\hline
    {\bf field gradient of}              & $\bullet$ {\bf charge confinement, ``optical tweezers''} \\
    {\bf standing wave}                  & $\bullet$ confinement of neutral particles in optical lattices \\ \hline
    field gradient of                    & $\bullet$ charged particle trapping\\
    moving wave                          & $\bullet$ laser wake field acceleration\\
    \hline
 \end{tabular}
 \caption{Examples of laser-matter interaction processes in macroscopic systems (for details and references see main text). 
          The concepts presently being used in dusty plasmas 
          are highlighted in bold, cf. also table~\ref{tab:clusters-dust}.
          Note that in many applications several laser properties are active simultaneously and the highlighted property should only be understood as the 
          dominant one. \label{tab:laser-matter}}
\end{table}

Traditionally, lasers have been used to excite electronic transitions in atoms and molecules allowing for high precision spectroscopy. By choosing a particular laser polarization, certain transitions can be activated or deactivated depending on the selection rules. In case of laser excitation of electrons into the continuum (photoionization) the photoelectron spectrum reveals detailed information on the target material. Related methods such as photoelectron spectroscopy are highly successful in atomic, molecular and condensed matter physics. 
With the increase of laser intensity also nonlinear processes such as multi-photon excitation and ionization have become available. In complex plasmas containing reactive species spectroscopic methods have found broad application as well, e.g. \cite{meichsner_recent_2012,bonitz_introduction_2010} and references therein, but this  has not been relevant for dusty plasmas so far. 
At the same time, ionization by UV radiation has been discussed by some groups~\cite{land_manipulating_2007, rosenberg_uv-induced_1995, rosenberg_uv-induced_1996, samarian_positively_2001, samarian_uv-induced_2000}, but has proven to be difficult. The interesting prediction is that the dust particles acquire a positive charge, but this will not be considered in the present paper.

Another important concept is the use of laser radiation with a spatially inhomogeneous intensity profile such as focused beams or interference patterns of several lasers. The latter gives rise to optical traps or optical lattices and has become a key tool in the field of ultracold atoms and molecules, e.g.~\cite{bloch_exploring_2005} or in the field of colloidal systems, where laser patterns have been used to create quasi-crystals \cite{Mikhael2011_SoftMatter,Bohlein2012_PRL}.
Furthermore, the spatially varying field of a moving laser wave is at the heart of particle acceleration predicted long ago 
\cite{dawson79, joshi84}. In the mean time laser wake field acceleration has been successfully realized experimentally 
and allows for the generation of relativistic electron beams with MeV energy \cite{pukhov02, katsouleas04, leemans04, ruhl06, lu_generating_2007} as well as for ion acceleration.
In dusty plasmas laser beams with spatially varying intensity profile are exploited utilizing the intensity dependence of the index of refraction of dust particles allowing to trap and move single particles (``laser tweezers'')~\cite{ashkin_observation_1986, ashkin_chapter_1997,ashkin_history_2000}, see also Ref. \cite{grier_revolution_2003}. 

An alternative direction in laser-matter interaction is heating of matter by generating Ohmic heat. Acceleration of electrons in solids easily allows to couple laser energy into the material and to melt it. This has become a standard method in technology and is used e.g. for microsopic structuring of surfaces as well as for drilling high quality micrometer-size holes. 
With the availability of ultrashort laser pulses it has become possible to couple the energy in a controlled way only into the electron subsystem, without heating the lattice, and to probe---on a femtosecond time scale---the ``non-thermal melting'' of solids far from equilibrium, e.g.~\cite{uschmann01,sokolowski03}. Lasers can also be directly used to heat plasmas, via photon absorption in electron-ion collisions (inverse bremsstrahlung), e.g.~\cite{kremp_PhysRevE.60.4725, bonitz_CTPP2150390407, haberland_PhysRevE.64.026405} and references therein. 
With the help of Petawatt lasers, the combination of laser ionization and particle acceleration allows to produce and compress plasmas which reach densities that are comparable to and even exceed by one or two orders that of metals. This has led to the new fields of laser plasmas or high energy density physics, e.g.~\cite{theobald_96, roth_fast_2001}. It has been predicted long ago that sufficiently high laser intensities will allow to achieve even fusion conditions \cite{tabak_94}.

The heating of dusty plasmas follows the same general idea of transmitting momentum and energy from the radiation to the particle ensemble by exploiting the light pressure but, obviously, using very modest laser intensities. The specifics here lies in the large particle size, compared to the typical laser focus, and in the peculiar properties of finite dust clusters. These mesoscopic particle ensembles have many properties in common with more traditional neutral or metal clusters and will be discussed below.

\subsection{Finite dust clusters and comparison with metal clusters}\label{ss:dust-metal}
The properties of finite systems have been first studied in nuclear matter and more recently in the context of cluster physics. Finite two-dimensional (2D) and spherical three-dimensional (3D) dust clusters in a (nearly) isotropic harmonic trap provide a fascinating opportunity to systematically study 
the physics of mesoscopic few-particle systems. The reason is the above mentioned unique opportunity provided by dusty plasmas to study structure, dynamics and thermodynamics of {\em all} individual system particles on the kinetic level, i.e. resolving simultaneously all particle trajectories. This is of high interest for finite systems in a variety of other fields including gas or metal clusters, electrons in quantum dots, trapped ions, ultracold gases and so on. In none of these fields observations on the kinetic level are possible.

Finite 2D clusters have been studied first, already for quite some time
\cite{juan_observation_1998,Lai99,Bedanov94,Schweigert95c,klindworth_laser-excited_2000,Melzer03}. More recently also
spherical 3D dust crystals could be produced \cite{Arp04}, and their low-temperature structure
(which, in general, is more sensitive to screening than for 2D systems) is now well understood (for details see
Sec.~\ref{ss:structure}).
In the case of strong coupling, the excitation behavior is dominated by collective modes of the whole system rather than by single particle dynamics.
This is typical not only for finite dust clusters but for mesoscopic systems in trapping potentials
in general. The frequency of some of these normal modes (in particular that of the monopole or
``breathing'' oscillation) has been found sensitive to the internal properties of finite systems
such as type of interaction \cite{henning_PhysRevLett.101.045002,olivetti_PhysRevLett.103.224301,kaehlert_pre14} and,
in the case of quantum clusters, also on the interaction strength (on the coupling parameter) \cite{bauch_PhysRevB.80.054515} or 
particle number \cite{abraham_1367-2630-16-1-013001}. This has led to the idea to use the normal
mode frequencies as a novel kind of ``spectroscopy'' for strongly correlated finite
systems~\cite{mcdonald_PhysRevLett.111.256801}, for a recent overview
see~\cite{abraham_CTPP:CTPP201300066}.

\begin{table}
 \begin{tabular}{|l||l|l|}
    \hline
    {\bf Property}	     & {\bf Metal clusters}             & {\bf Dust clusters}	  \\ \hline\hline
    size, geometry           & $N \sim 10 \dots 10^4$; 3D        & $N \sim 10 \dots 10^4$; 2D, 3D\\
\hline
    elementary constituents  & single atoms                      & spherical plastic particles\\
    particle radius $R$      & $R\sim a_B$                       & $R\sim 1 \dots 10 \mu m$\\
\hline
    interparticle distance ${\bar r}$  & ${\bar r} \sim $ few $a_B$ & ${\bar r}$ several 100 $\mu m$\\
\hline
    confinement              & attraction from                   & external potentials: \\
                             & central ionic core                & trap, thermophoresis etc.\\ 
\hline
    pair interaction         & Coulomb repulsion                 & screened Coulomb   \\ 
                             &  between electrons                & (Yukawa) repulsion \\  
\hline
    interaction strength     & fixed                             & externally controlled  \\ 
\hline
    coupling parameter       & valence electrons:                & dust component:  \\ 
                             & $r_s = {\bar r}/a_B \sim 2\dots 5$ & $\Gamma = E_{\rm int}/E_{\rm kin} \sim 10 \dots 10^4$ \\
\hline
    stationary state         & equilibrium                       & non-equilibrium$^a$ \\
\hline  
    ground state             & concentric shells                 & concentric shells      \\
\hline\hline
    {\bf Laser manipulation} & $\sigma \gg R, {\bar r}$          & ${\bar r} \gtrsim \sigma > R$ \\
    laser spot size $\sigma$ & homogeneous field                 & fast spot movement: \\
                             & across cluster                    & on average homogeneous \\
                             &                                   & force across cluster\\
                             & $\bullet$ photoionization,        & $\bullet$ single-particle diagnostics\\
                             & $\bullet$ collective ionization   & $\quad$ by light scattering                 \\ 
                             & $\bullet$ Coulomb explosion       & $\bullet$ particle control \\
                             &                                   & $\quad$ by optical tweezers \\
    laser heating            & $\bullet$ via inv. bremsstrahlung & $\bullet$ via light pressure \\ \hline
 \end{tabular}
 \caption{Comparison of atomic clusters (example of metal clusters) with finite dust clusters. In the lower part the relevant cluster 
           manipulation approaches by means of lasers are outlined. Comments: $a$--streaming electrons and ions may give rise to 
           anisotropic and even attractive dust-dust interaction, depending on the plasma parameters and location in the discharge, 
           e.g. \cite{kroll2010,ludwig_njp,block_cpp12}.
           \label{tab:clusters-dust}}
\end{table}

Of particular interest is the gradual crossover, with increased cluster size~\cite{juan_observation_1998, klindworth_laser-excited_2000}, from single atoms to macroscopic condensed matter~\cite{meiwes-broer_metal_2000, haberland_melting_2005}. In this field, also the interaction of lasers with finite clusters has been studied in great detail. Due to the similarity of these systems to the finite dust clusters that are in the focus of this review we provide some comparison in table~\ref{tab:clusters-dust}. In both cases clusters in a similar range of particle number are studied. The main difference is the different type of particles---atoms (or molecules) versus highly charged plastic spheres---and the different type of confinement. In the case of metal clusters the valence electrons are kept together by the central Coulomb force from the ionic core of the cluster. In contrast, in the case of dust clusters the pair interactions are purely repulsive and the confinement is provided by an external potential (or combinations thereof). Another key difference is that, in metal clusters, the coupling strength is fixed by the governing Coulomb forces between all particles. In contrast, in dust clusters the charge of the grains and their distance can be modified by changing the plasma parameters. Nevertheless, the dust systems usually feature only relatively large values of the coupling parameter $\Gamma$.

Among the most interesting questions both for finite metal clusters and finite dust clusters is the size dependence of their properties. This includes the ground state structure, the excitation spectrum and the thermodynamic properties. Since a central topic of this review is phase transitions this involves the question of size dependent melting temperatures. It is known for a long time that the melting temperature of small clusters is lower than the bulk melting temperature. This topic has been studied for metal clusters in some detail. We mention experimental studies, e.g. \cite{buffat_PhysRevA.13.2287,lai_apl98,dippel_PhysRevLett.87.095505} and references therein, theoretical work, e.g. \cite{reiss_jcp88,Vanfleet199540} as well as computer simulations, e.g. \cite{qi_jcp01, ding_PhysRevB.70.075416}. There is overall consensus that the melting temperature of such clusters with short-range interaction decreases proportional with the cluster diameter. For harmonically confined Coulomb clusters \cite{schiffer_melting_2002} also a decrease of the melting temperature was observed in simulations which is almost linear in the fraction of particles in the surface layer. For dust clusters, no such general trend is known yet due to the lack of systematic studies of larger clusters and due to the difficulties in controlled cluster heating. The laser heating technique described in this review will pave the way towards such studies.

When lasers where first shot on finite metal clusters, the experiments revealed a surprisingly effective energy absorption \cite{ditmire_PhysRevLett.78.3121}. Subsequently, also collective absorption mechanisms (by coupling to plasmons) were observed \cite{meiwes_PhysRevLett.82.3783,milchberg_pop05}. These effects could be successfully explained by theoretical modeling \cite{milchberg_PhysRevE.64.056402} and complex many-particle simulations, for an overview see \cite{fennel_RevModPhys.82.1793}.
Among the theoretical difficulties is the proper treatment of electronic quantum many-body effects, see e.g. \cite{dufty_jpcs05,Fromm20083158}. A detailed discussion of the various laser-cluster interaction mechanisms has been given in Ref.~\cite{rost_0953-4075-39-4-R01}, see also the recent reviews \cite{Krainov2002237,fennel_RevModPhys.82.1793}.

Among the most interesting effects of laser illumination of metal clusters is the emission of higher harmonics of the laser radiation \cite{ditmire_harmonics_PhysRevLett.98.123902}, the emission of energetic electrons and, in case of strong ionization, Coulomb explosion of the whole cluster \cite{ditmire_PhysRevA.83.043201}.
A similar rapid expansion of Coulomb and Yukawa dust clusters has been recently predicted in Ref.~\cite{piel_13}. 
Here laser illumination is not necessary at all, it is sufficient to turn off the confinement potential. 
Similarly, many other effects observed for metal clusters in the presence of intense laser radiation can also be expected for dust clusters. 
However, more recently the focus was on a gentle heating of dust clusters so that the system is transfered from its ground state into a state with moderately elevated temperature.

As discussed above a focused low intensity laser beam is well capable to accelerate single dust particles. 
Similarly, stationary laser beams were used to excite shear flows in monolayer dust crystals~\cite{nosenko_shear_2004,feng_observation_2012} or rotations in finite 2D clusters~\cite{klindworth_laser-excited_2000}. 
Mach cones could be excited by moving the laser spot through the dust crystal~\cite{melzer_laser-excited_2000,nosenko_observation_2002}. 
A further application of moving laser spots is the realization of a heat source for the dust component~\cite{wolter_laser_2005, nosenko_laser_2006, nosenko_heat_2008,Goree13}.
A more detailed description of these pioneering laser experiments is given in section~\ref{sec:general_concept}.
An improved heating scheme allowing to realize a true thermodynamic heating with an isotropic, Maxwellian 2D velocity distribution has been applied by Schablinski \textit{et al.}~\cite{schablinski_laser_2012} to finite 2D dust clusters. 
A first heating concept for finite 3D clusters has been realized by Schella \textit{et al.}~\cite{Schella11}, see Sec.~\ref{sec:2Dexp}.

The goal of this review article is to present an overview on these recent experimental
developments, compare them to theory and computer simulations and to discuss possible future applications of laser manipulation of dust clusters. We start by
giving a brief overview on the properties of finite 2D and 3D dust clusters in
Sec.~\ref{s:td_trans}. Then, we discuss the laser heating principle and how it is used to reduce
the coupling strength in a controlled way (Sec.~\ref{s:control}).
Dedicated numerical simulations of the laser heating are presented in Sec.~\ref{s:numerics}.
Experimental results for finite 2D and 3D dust systems are presented in
Secs.~\ref{sec:2Dexp} and \ref{s:3d}. We conclude in Sec.~\ref{s:discus} with an outline of future
applications, including spatially inhomogeneous plasmas and time-dependent processes.

\section{Structural, thermodynamic and transport properties \\of finite dust clusters}\label{s:td_trans}

\subsection{Structural properties of extended dust clusters}

As described in the introduction, electrons and ions screen the repulsive interaction of the dust grains.
The interaction between the dust particles is, in most cases, well described by a Yukawa-type
pair-potential
\begin{equation}
 \Phi_{\rm Y} \left(\vec{r}_i,\vec{r}_j\right) = \frac{Q_{\rm d}^2}{4\pi \varepsilon_0\, \left|\vec{r}_i - \vec{r}_j \right|}\cdot e^{-\kappa \left|\vec{r}_i - \vec{r}_j \right|} \ ,
 \label{eq:Yukawa}
\end{equation}
where $\kappa=\lambda_{\rm D}^{-1}$ is given by the inverse Debye length taking into account
the screening effect of electrons and ions\footnote{The Debye length $\lambda_{\rm
D}=\left(\frac{q_{\rm e}^2 \bar{n}_{\rm e}}{\varepsilon_0 \kB T_{\rm e}} + \frac{q_{\rm i}^2
\bar{n}_{\rm i}}{\varepsilon_0 \kB T_{\rm i}} \right)^{-1/2}$ incorporates screening of electrons
and ions. $q_{\rm e}$ ($q_{\rm i}$) is the electron (ion) charge, $\bar{n}_e$ ($\bar{n}_i$) is the
average electron (ion) density and $T_{\rm e}$ ($T_{\rm i}$) is the electron (ion) temperature.}.
Wake effects due to the streaming ions (for recent overviews see e.g. Refs.~\cite{kroll2010,ludwig_njp,block_cpp12}) are neglected in the presented results.

Extended 2D dust systems arrange, in the solid state, in a hexagonal lattice~\cite{Chu94b,thomas_plasma_1994,Hayashi94}.
These systems are of special interest since the mechanism of  thermodynamic phase transition in 2D systems from the ordered solid phase to the unordered liquid phase is still not finally clarified~\cite{Strandburg88}.
On the one hand, a first-order transition is predicted by the formation of grain boundaries between crystalline patches~\cite{Chui83,Strandburg88}.
On the other hand, a two-step second-order transition with an intermediate hexatic phase, the so-called KTHNY scenario~\cite{Kosterlitz73,Nelson79,Young79,Strandburg88}, is expected.
A recent numerical study concludes that the hexatic phase is metastable and vanishes in the long-time limit~\cite{hartmann_hexatic14}.
Thermodynamic heating of 2D dust crystals by means of laser techniques might, therefore, be very beneficial in addressing these issues from the experimental side, see Sec.~\ref{sec:2Dexp_ext}.

\subsection{Structural properties of finite dust clusters}\label{ss:structure}
The striking property of solid Yukawa clusters in 2D as well as in 3D is their well ordered
structure. This structure and the loss of order with increasing temperature is accompanied by a
sequence of phase transitions (or structural transitions) which are peculiar in finite
systems~\cite{Bedanov94,apolinario_melting_2007}. Details of these transitions are still open
and of high interest for many finite size systems. They depend on the particular crystal structure
which we, therefore, review in the following.

The Hamiltonian of the $N$-particle Yukawa cluster is (prior to laser manipulation) given by~\cite{bonitz_introduction_2010}
\begin{equation}
 \mathcal{H} = \sum_{i=1}^{N} 	\frac{1}{2m} \vec{p}_{i}^{2}
	      +\sum_{i=1}^{N} 	\frac{m\omega_0^2}{2} \vec{r}_{i}^{2}
	      +\sum_{i<j}	\Phi_{\rm Y} \left(\vec{r}_i,\vec{r}_j\right) \ ,
  \label{eq:dust_hamilton}
\end{equation}
where the first term describes the kinetic energy, the second the confinement energy due to the
harmonic trap of strength $\omega_0$ and the third the mutual Yukawa interaction energy. In this model, all dust grains
are assumed to be equal in mass $m$ and charge $Q_{\rm d}$.

Small 2D dust clusters typically consist of concentric rings~\cite{juan_observation_1998,Bedanov94}.
The core region of larger clusters with several hundred particles, in contrast, shows a hexagonal structure, like in infinite systems.
This lattice has dislocations at the outer shells, where the circular boundary has to be matched~\cite{kong_topological_2003}.

Finite 3D dust clusters consist of spherical shells instead of rings~\cite{Arp04,Arp05}.
Due to their spherical shape, these clusters are called \textit{Yukawa balls} [or \textit{Coulomb balls}, when screening can be neglected]. 
This structure has been measured experimentally~\cite{Arp04,block_structural_2007} and is reproduced in first-principle simulations--molecular dynamics (MD) or Monte Carlo (MC)--and has been investigated in detail for both Coulomb balls~\cite{hasse_structure_1991,schiffer_layered_1988,ludwig_structure_2005} and Yukawa balls~\cite{bonitz_structural_2006, baumgartner_ground_2008}. 
The general trend is that, when the screening parameter $\kappa$ increases, more particles occupy the inner shells, and--even in the ground state--the average density becomes inhomogeneous, decaying towards the cluster boundary~\cite{henning_ground_2006,henning_ground_2007}.
This is in striking contrast to classical Coulomb clusters which have, at $T=0$, a homogeneous mean density.\footnote{Nevertheless, as a reasonable first approximation 
for the average density one can often use the particle number $N$ and the radius of the outermost shell $R_{\rm C}$, $\langle n \rangle \approx N/(4\pi R^3_{\rm C}/3)$}

Recently, also several analytical theories for the shell structure in 3D have been developed. The
\textit{local density approximation} as a continuum theory accurately describes the mean
density profile of the spherical clusters~\cite{henning_ground_2006,henning_ground_2007},
but it misses the formation of shells. The positions and the populations of the shells of
Coulomb balls are well reproduced by a slightly modified version of the \textit{hypernetted
chain approximation} which can be adapted to particles interacting via a Yukawa potential as
well~\cite{wrighton_theoretical_2009,wrighton_shell_2010,bruhn_theoretical_2011,wrighton_charge_2012}.
Beyond the radial shell structure, 3D dust balls exhibit a well ordered intra-shell structure at strong coupling.
In contrast to a flat 2D system, the spherical curvature requires a fraction of pentagonal Voronoi cells in the hexagonal pattern on the shell~\cite{ludwig_structure_2005,kading_three-dimensional_2006}.
\begin{figure}
 \includegraphics{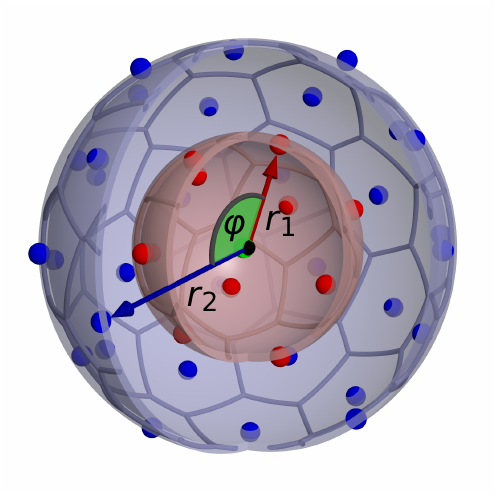}
 \caption{Yukawa ball with $N=60$ particles consisting of two spherical shells and a single particle in the center. In order to sample the center-two-particle (C2P) correlation function based on the coordinates, for each particle pair, both radial coordinates $\rone$ and $\rtwo$ as well as the angular pair distance $\gangle$ with respect to the trap center are recorded. The Voronoi grid of the intra-shell particle configuration is shown in gray.\label{fig:c2pexplain}}
\end{figure}
\subsection{Thermodynamic properties and phase transitions}\label{ss:tdyn}
As the ground state properties of finite dust clusters are well understood by
now~\cite{hasse_structure_1991, schiffer_layered_1988, ludwig_structure_2005,
bonitz_structural_2006, baumgartner_ground_2008}, further investigations concentrate on
thermodynamic properties at elevated temperatures. When a cluster is excited by feeding thermal
energy into the system, metastable states with energies above the ground state energy
$E_0(N,\kappa)$ are occupied. These metastable states may differ from the ground state with
respect to the occupation numbers of the shells or in the particle configuration within the
shells. The metastable states of Yukawa balls as well as their increased population with
temperature were investigated in both experiment~\cite{Kaeding08,block_structural_2007} and
first principle simulations~\cite{kaehlert_pre08,baumgartner_structure_2009}. Among others,
these simulations allowed to determine the heights of energy barriers between different metastable
states~\cite{golub_jpa06,kaehlert_pre08,baumgartner_structure_2009}. This ``fine structure''~\cite{ludwig_structure_2005} is observed for 3D but not for 2D clusters. 

When the temperature is increased a transition from a well ordered structure with thin shells and a highly symmetric intra-shell
order (3D) towards a disordered particle fluid-like state is observed. It is an interesting question whether this 
process occurs rapid and constitutes a phase transition, as in macrocscopic systems. While many similarities to phase transitions have indeed been observed\footnote{As the solid and fluid ``phases'' are, strictly speaking, concepts defined in the macroscopic limit the same holds true with phase transitions. Nevertheless, the processes in many finite systems are very similar and the application of thermodynamic concepts is useful. Often though the term ``crossover'' is preferred in comparison to phase transition.},
there are also differences: the first is that in finite systems ``melting'' requires a finite temperature interval.
The second is that melting may involve a sequence of distinct processes. Therefore, in order to resolve and understand these processes, 
the concepts for characterization of phase transitions known from macroscopic systems have to be re-considered and adapted.
In particular, structural parameters are required that are suitable to characterize the phase transition. The first quantity that comes to mind is the heat
capacity, (the amount of heat $\delta Q$ required  to heat up the system a temperature $\delta T$) which is a widely used melting parameter e.g. in solid state physics. However, measuring the heat capacity is challenging in dusty plasmas. The dust subsystem exchanges energy with
electrons, ions and neutrals making it very difficult to extract the pure heat capacity of the dust
system with a particle number that is negligible compared to the number of the
surrounding plasma constituents~\cite{schiffer_melting_2002}. In contrast, the pair distribution
function and, even more,  the center-two-particle correlation, discussed 
in the next section, have proven to be well suited for this purpose in theory as well
as in experiments.

Let us now summarize the known results about ``phase transitions'' in finite dust systems.
In two dimensions, for the loss of the ring structure, two different melting processes were identified. The first
process is attributed to the rotation of one ring with respect to the other
rings~\cite{Bedanov94,Schweigert95c,klindworth_laser-excited_2000}. The required energy for
such a rotation crucially depends on the exact occupation number of the rings. For example, the
commensurate configuration (``magic number'') $N=19$ (1-6-12) is very stable against such
inter-shell rotations (due to the matching particle numbers on the inner and outer shell this cluster has a perfect hexagonal symmetry). In contrast, the
$N=20$ (1-7-12) configuration is extremely unstable against this
excitation~\cite{klindworth_laser-excited_2000} and has a drastically reduced inter-shell
rotation barrier and melting temperature. The second melting process is associated with
particles undergoing a transition between two adjacent rings and is called ``radial melting''. It, typically, takes place at
substantially higher temperatures~\cite{ivanov_melting_2005}.
It is worth noting that the same kind of two-stage melting process is observed in finite quantum clusters \cite{filinov_pss00, filinov_prl01}, indicating that these are correlation effects which are of high interest also beyond the field of dusty plasmas.

The complexity of the melting process increases when advancing from 2D to 3D clusters
\cite{golub_jpa06, Apolinario07}. Besides the melting of the radial structure and the inter-shell
rotation, a third melting process emerges that is connected to the intra-shell order~\cite{Apolinario07}.
However, this classification is not strict since the interplay between the different melting
processes in 3D clusters is utmost complex and there is, in general, no separation of the different processes. Thus many interesting questions are still open that have to be answered by experiments and theory.

In order to trigger phase transitions in dusty plasma experiments, selective control over the dust kinetic
temperature is essential. In particular, it is desirable to feed energy into the random dust motion
without changing other plasma parameters such as the neutral gas pressure, the electron and
ion temperatures and the flow velocity of the ions. A further requirement is that the entire
cluster should be heated homogeneously while preserving an isotropic velocity distribution.
As will be described in detail in Section~\ref{s:control}, this selective control over the dust temperature is
indeed possible---by the means of intense laser light.

\begin{figure}
 \includegraphics{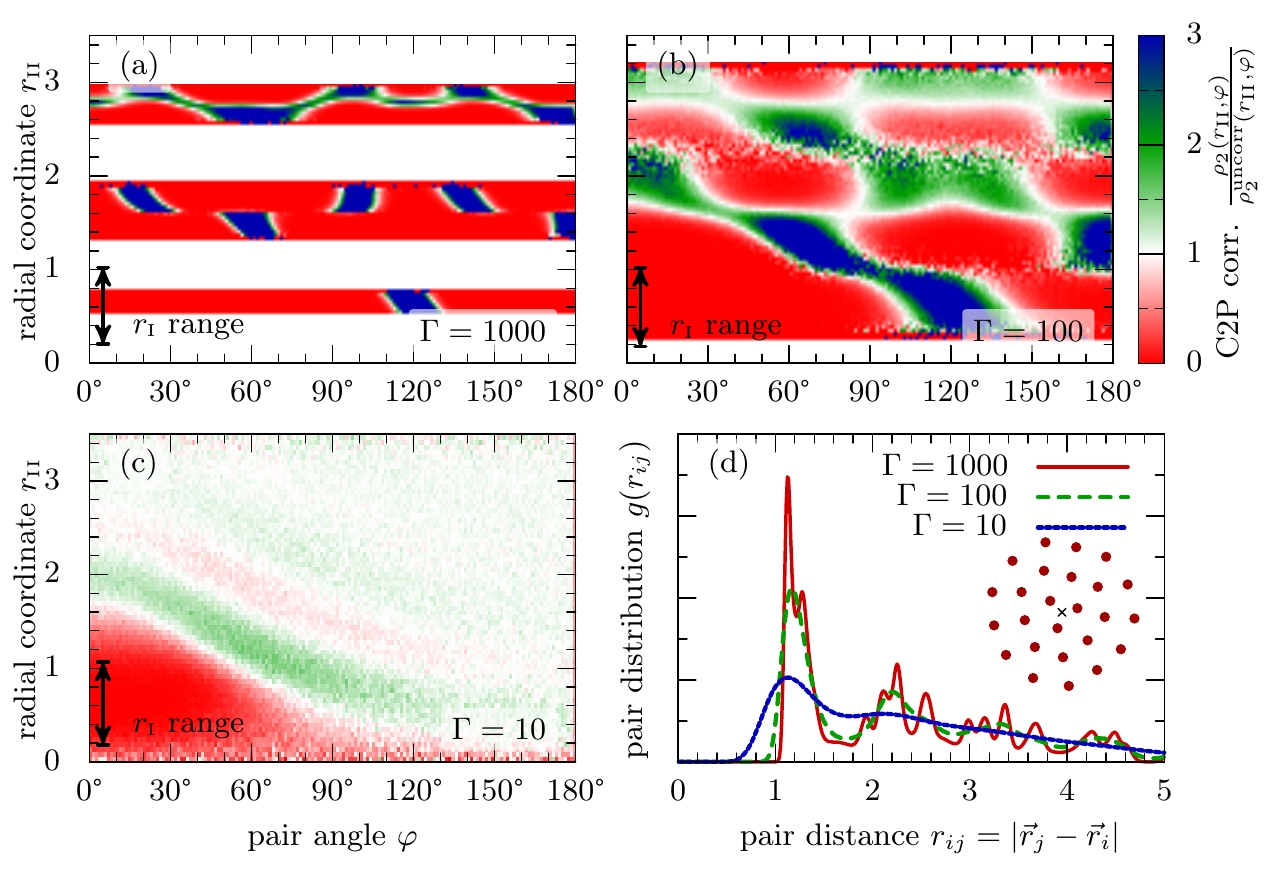}
 \caption{(a)-(c) Center-two-particle correlation function for a 2D Coulomb
cluster with $N = 25$ particles for different coupling strengths. The ground state
configuration consists of three concentric rings, see inset in (d). The
first radial coordinate $\rone$ is integrated over a range corresponding to the inner
shell indicated by the arrows. Intra-shell correlations are visible at $\rtwo \sim 0.7$ and $\phi = 120$.
Pronounced inter-shell correlations with particles on the second shell are found
at $\rone \sim 1.6$ for strong coupling. (d) The radial pair distribution function $g(r_{ij})$
shows distinct peaks at high $\Gamma$ which vanish when $\Gamma$ is being decreased. Note that $g(r_{ij})$
decreases to zero at large distances for all couplings, due to the finite size of the cluster.
Results are from a Monte Carlo simulation, $\Gamma$ is defined with the length unit
$r_0 =[Q_d /(4\pi\epsilon_0 m \omega_0)]^{1/3}$ as characteristic pair distance.
\label{fig:c2p_2d}}
\end{figure}
\subsection{Key quantities for the analysis of finite dust clusters}\label{ss:quantities}
An important structural parameter that characterizes the order (and its loss during melting) in extended systems is the
\textit{radial pair distribution function} $g(r_{ij})=g\left(\left|\vec{r}_i - \vec{r}_j \right|\right)$. It
is commonly defined by the average number of particle pairs found at a distance of $r_{ij}$
divided by the number of pairs which one would find in a homogeneous (i.e. uncorrelated)
system with the same density. The first maximum of $g$ reflects the mean inter-particle distance between the particles.
An algebraic decay of the envelope of $g$ over $r_{ij}$ indicates a
long range order, characteristic of the solid regime. Moreover, the height of the first maximum allows to detect
a melting line in the $(\Gamma,\kappa)$ space~\cite{ott_non-invasive_2011}. 

\begin{figure}
 \includegraphics{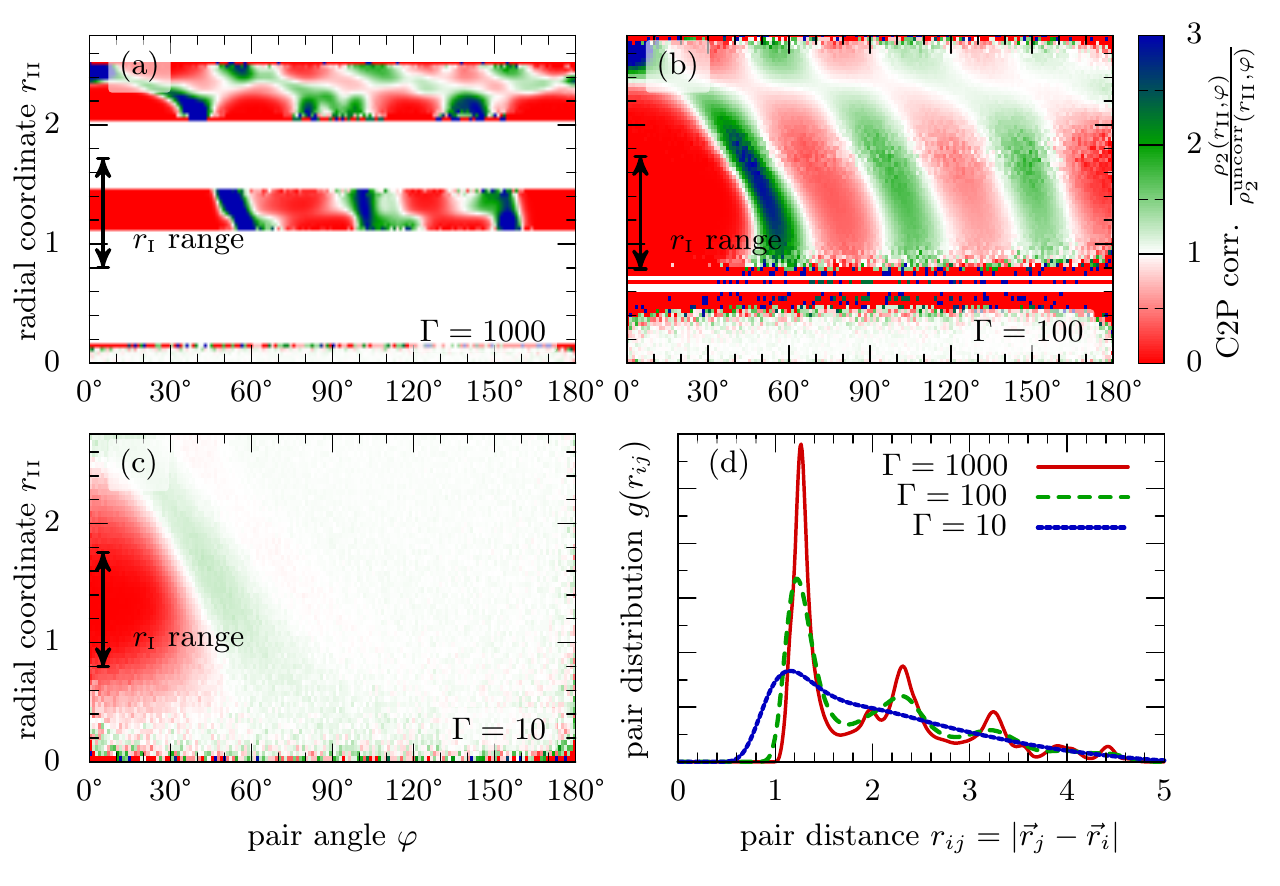}
 \caption{(a)-(c) Center-two-particle correlation function for a 3D Yukawa ($\kappa=1$) ball 
  with $N = 60$ particles for different coupling strengths. The ground state
  configuration consists of two concentric spherical shells and one particle in the center. 
  The first radial coordinate $\rone$ is integrated over a range corresponding to the inner
shell (arrows). {\em Intra-shell} correlations are visible at $\rtwo \sim 1.3$.
{\em Inter-shell} correlations with particles on the outer shell are visible
at $\rone \sim 2.2$ by the angular modulation of $g_2(\rtwo,\gangle)$. 
(d) The radial pair distribution function $g(r_{ij})$ shows distinct peaks a high $\Gamma$ which vanish when $\Gamma$ is being decreased.
Note that $g(r_{ij})$ decreases to zero a large distances for all couplings, due to the finite size of the cluster.
Results are from a Monte Carlo simulations, $\Gamma$ is defined with the length unit $r_0 =[Q_d /(4\pi\epsilon_0 m \omega_0)]^{1/3}$ as characteristic pair distance.
\label{fig:c2p_3d}}
\end{figure}

The radial pair distribution function for finite 2D and 3D clusters is shown in parts d of~Figs.~\ref{fig:c2p_2d} and
~\ref{fig:c2p_3d}, respectively\footnote{The pair distribution function of finite clusters is computed by introducing discrete bins of width $\Delta r$, counting the average number of particle pairs with a distance in the interval $\left[r_{ij},r_{ij}+\Delta r\right)$ and dividing by the bin volume.}. $g(r)$ drops to zero after a few nearest
neighbor distances for finite clusters in both 2D and 3D. For obvious
reason, a true ``long range behavior'' cannot be investigated for finite
clusters. Nevertheless, $g(r)$ contains detailed information about the thermodynamic state of the cluster and its temperature dependence.
In particular, a fine structure of the peaks in $g(r_{ij})$ is visible
at high coupling strength (low temperature), indicating a frozen structure. When the temperature is increased, the
subpeaks disappear. Finally, at a moderate coupling strength, see curve $\Gamma=10$ in Fig.~\ref{fig:c2p_2d}, the pair
distribution function has only a single peak followed by a monotonous decay. The classification
of inter-shell and intra-shell melting is not possible by means of this quantity as sampling the
modulus of the distance does no distinguish whether the particles of a pair are on one shell or
on different shells.

For this reason, the \textit{center-two-particle} (C2P) correlation function $g_2 \left(\rone, \rtwo, \gangle \right)$ is introduced which takes into account the radial position of both particles as well as their angular distance with respect to the trap center~\cite{ludwig_tuning_2010, Schella11}.
A sketch of the sampled coordinates is shown in Fig.~\ref{fig:c2pexplain} for a 3D Yukawa ball consisting of two spherical shells and one particle in the center.
To evaluate the C2P, the sampled two-particle probability density $\rho_2 \left(\rone, \rtwo, \gangle \right)$ is computed and normalized by the uncorrelated probability density (i.e. the function which one would find in a system with the same radial structure but homogeneously filled shells) $\rho_2^{\rm uncorr}\left(\rone, \rtwo, \gangle \right)$,
\begin{equation}
  g_2\left(\rone, \rtwo, \gangle \right) = 
    \frac{\rho_2\left(\rone, \rtwo, \gangle \right)}
      {\rho_2^{\rm uncorr}\left(\rone, \rtwo, \gangle \right)} {\rm .}
 \label{eq:c2p_correlation}
\end{equation}
Since $\rho_2 \left(\rone, \rtwo, \gangle \right)$ is invariant under
rotation of the entire cluster, it is not necessary to remove a rigid rotation before sampling.

The C2P contains very detailed information on the structure of finite clusters and useful special cases. For example,
integration of $g_2$ over both radial coordinates over a range corresponding to one shell
allows one to extract the angular pair correlation function within that shell. On the other hand,
when only one radius coordinate is integrated of the width of one shell, the TCP contains information on the relative radial and angular arrangement of 
the particles of the same shell as well as from different shells. To visualize this,
$g_2 \left( \rone, \gangle \right)$ can be plotted in a color map, as
is done for a 2D cluster, in Fig.~\ref{fig:c2p_2d}, and a 3D cluster, cf. Fig.~\ref{fig:c2p_3d}. In
this color map, the intra-shell structure is responsible for the maxima and minima in $\gangle$-direction.
While the data in the figures are from MC simulations, equally well one can use
 data from experiments with finite dust clusters, as is shown below in Fig.~\ref{fig:TCF}.

For the 2D cluster with $N=25$ particles, depicted in Fig.~\ref{fig:c2p_2d}, a reference particle from the inner shell is chosen by
the integration range around $\rone \approx 0.7$. Since the inner shell consists of three
particles, the intra-shell neighbors appear as a peak at $\gangle=120^\circ$. At
$\Gamma=1000$, the distinct peaks found around $\rtwo \approx 1.6$ show that the angular
orientation of the second shell with nine particles is locked with respect to the inner shell. This
inter-shell order disappears between $\Gamma=1000$ and $\Gamma=100$ where $g_2$ is
smeared out in angular direction. At moderate coupling, $\Gamma=10$, hardly any angular
correlations remain and also the radial order is lost. Particle transitions between different shells
are revealed by a finite density in the radial regions between the shells.\footnote{Note that, at very strong coupling, the correlation function in the region between two shells cannot be calculated since both $\rho_2$ and $\rho_2^{\rm uncorr}$ become zero.}

Consider now a 3D Yukawa cluster with $N=60$ particles, cf. Fig.~\ref{fig:c2p_3d}. 
In the ground state (and at strong coupling), the  particles are found in a configuration
which has one particle in the center, 15 particles on the inner shell and 44 particles on the outer
shell (44-15-1). Again, one reference particle is chosen from the inner shell by integrating over
$\rone$. Due to the complexity of the particle composition on a spherical shell compared to the
composition on a ring in 2D, the peaks at $\rtwo \approx 1.2$, indicating intra-shell
correlations, are not as sharp as for the 2D cluster at high coupling. Inter-shell correlations
appear as dark and bright areas in the horizontal stripe at radius $\rtwo \approx 2.4$ that
corresponds to the outer shell. At moderate coupling, $\Gamma=10$, the angular correlations
are lost and frequent transitions between the shells take place as seen by the radial extension of
the density.
\subsection{Transport properties}\label{ss:transport}
Besides thermodynamic properties, also transport coefficients and their dependence on parameters like temperature or the magnetic field strength are important characteristics of dusty plasmas.
A particularly important example of transport coefficients is the diffusion coefficient.
In dusty plasmas, diffusion was investigated in detail in macroscopic systems, e.g. \cite{daligault_liquid-state_2006,PhysRevLett.107.135003} and references therein.
In particular, in two dimensions an anomalous diffusion was found~\cite{juan_anomalous_1998, PhysRevLett.102.085002, ott_prl09, ott_cpp09} which turned out to be a transient effect \cite{PhysRevLett.103.195001}.
Furthermore simulations were performed for magnetized dusty plasmas. Here, the diffusion coefficient
$D_\perp$ perpendicular to the magnetic field as well as the parallel diffusion coefficient
$D_\parallel$ were found to be strongly affected by the magnetic field in the strong coupling
regime approaching Bohmian diffusion (decay with $B^{-1}$)~\cite{PhysRevLett.107.135003}.
Recently diffusion in a two-dimensional one- and two-component magnetized strongly coupled
plasma was studied and interesting behavior of the diffusion coefficients of both components
was reported \cite{ott_pre14}.

One way to compute (or measure) the diffusion coefficient in a macroscopic system is to use the
mean square displacement (MSD)
\begin{equation}
 u_r(t) = \left\langle \left| \vec{r}(t) - \vec{r}(t_0) \right|^2 \right\rangle_{N} = 2 \,{\dim}\, D t^{\alpha} {\rm ,}
 \label{eq:mean-square-displacement}
\end{equation}
as an average over all $N$ particles. Here dim is the system dimensionality (2 or 3) and
$\alpha$ is the diffusion exponent which equals one for normal diffusion. However, the long
time behavior of the MSD has only limited meaning for mesoscopic systems since there the
particles reach the cluster border after a few inter-particle
distances~\cite{vaulina_evolution_2008}.

Therefore, a different method is required to calculate the diffusion coefficients in small clusters.
The instantaneous normal mode (INM) analysis has proven successful to this
purpose~~\cite{Seeley89,Stratt95, Keyes97,Vijayadamodar95,Keyes96,Melzer13}. This method
deduces dynamic properties of a liquid state from the curvature of the energy landscape of the
momentary configuration of the cluster. The first step is to calculate the dynamical (Hessian)
matrix of the potential energy in the Hamiltonian, Eq.~(\ref{eq:dust_hamilton}), as
\begin{equation}
 \mathbf{A} = \left(\frac{\partial^2 E}{\partial r_{i\alpha} \partial r_{j\beta}} \right) {\rm ,}
 \label{eq:dynamical-matrix}
\end{equation}
where $i,j$ are particle indices and $\alpha,\beta$ indicate the (two or three) components of the
coordinates.
The eigenvalues of the this matrix present the squared eigenfrequencies $\omega_l^2$ of the system. 
In a stable state, all these eigenvalues are positive resulting in real eigenfrequencies which reflect stable oscillations of the particles in the potential cage formed by their neighbors~\cite{donko_caging_2002, Das04}.
A liquid instantaneous configuration, in contrast, has also negative eigenvalues of $\mathbf{A}$, resulting in purely imaginary eigenfrequencies reflecting unstable modes.

The second step is to calculate the spectral density, $\rho(\omega)$, which is normalized to unity, $\int{{\rm d}\omega \rho(\omega)}=1$.
In turn, we average over the normal modes of many configurations
\begin{equation}
 \rho(\omega) = \left\langle \sum_{l=1}^{{\rm dim}\cdot N}{\delta(\omega - \omega_l)} \right\rangle {\rm .}
 \label{eq:density-of-states}
\end{equation}
This density is composed of a stable part, $\rho_{\rm s}(\omega)$, with real frequencies, and an unstable part, $\rho_{\rm u}(\omega)$, with imaginary frequencies.
The unstable part $\rho_{\rm u}(\omega)$ is associated to a negative curvature in the momentary potential landscape.
As described in Refs.~\cite{Seeley89,Stratt95, Keyes97,Vijayadamodar95,Keyes96,Melzer13}, especially $\rho_{\rm u}(\omega)$ can be related to the diffusion constant.
The self-diffusion constant is expressed as~\cite{Seeley89,Stratt95, Keyes97,Vijayadamodar95,Keyes96,Melzer13}
\begin{equation}
 D = \frac{\kB T}{m} \int{\rho(\omega) \frac{\tau_{\rm h}}{1+\tau_{\rm h}^2 \omega^2} {\rm d}\omega} {\rm ,}
 \label{eq:diffusion-omega-integral}
\end{equation}
and depends on the temperature $\kB T$, the particle mass $m$ and the average ``hopping time''
$\tau_{\rm h}$ for the transition across potential barriers between two local potential wells.
This time is known as the inverse hopping frequency and calculated as
\begin{equation}
 \tau_{\rm h}^{-1} = c \int{\frac{\omega}{2\pi} \rho(\omega) A \exp\left(-B \frac{\omega^2}{\kB T} \right) {\rm d}\omega} {\rm ,}
 \label{eq:hopping-frequency}
\end{equation}
where $c \approx 3$ is a constant taking into account the different routes to escape from a local
potential minimum and the constants $A$ and $B$ are obtained from an exponential fit of
$\rho_{\rm u}(|\omega|) / \rho_{\rm s}(\omega)$. For details, we refer the reader to
Refs.~\cite{Seeley89,Stratt95, Keyes97,Vijayadamodar95,Keyes96,Melzer13}.

\section{Controlled change of the coupling strength by laser manipulation}\label{s:control}

\subsection{General concept\label{sec:general_concept}}
In order to obtain valid information on thermodynamic properties it is essential to gain reliable
control on the coupling parameter of the system. According to Eq.~(\ref{eq:Coulomb_Gamma}),
three possibilities exist to control $\Gamma$: (i) controlling the charge $Q_{\rm d}$ of the
particles, (ii) controlling the inter-particle distance $b_{\rm WS}$, and (iii) controlling
temperature $T$.
With $\Gamma \sim Q_{\rm d}^2$, already a moderate variation of charge allows to change the
coupling strength considerably. Thus, controlling the charge on the particles is the most
tempting approach. However, this is not feasible in practice. $Q_{\rm d}$ is determined by
geometric properties of the particles (such as the particle radius) and by the plasma conditions in the vicinity of the
particles. Except for rare special conditions, the geometric properties cannot be altered during
experiments and the local plasma conditions are not solely set by the discharge parameters but
are modified by neighboring particles as well. Thus, the charge is not directly controllable by
means of external parameters such as discharge power, neutral gas pressure or bias voltages.
Each will affect the plasma as well as the particle arrangement and thus result in a rather
complex parameter dependence. Especially the inter-particle distance is strongly determined by
the mutual particle repulsion, i.e. the particle charge, and the external confinement. Thus, there
is no easy access to control charge and inter-particle distance independently.
Therefore, to decrease the coupling strength, temperature is the only remaining control parameter.
\subsection{Kinetic and surface temperature of dust particles}\label{ss:t_ts}
Due to the macroscopic size of the particles (at least) two temperatures have to be distinguished: the kinetic temperature $T^i$ of the particles and the particle surface temperature $T^i_s$, where ``i'' labels different dust particles (we use the notion ``temperature'' for the mean kinetic energy of the particle). To understand the physical difference of these temperatures we have to take a microscopic view point: The dust grains are solid bodies consisting of a very large number  $N$ of molecules (their atomic substructure and electronic properties are not relevant in this context). A transfer of energy to the particle from external sources will, in general, excite all $3N$ degrees of freedom of the molecules. Among them are $6$ degrees of freedom (``center of mass'', COM modes) which are related to a collective displacement or a rotation of all particles which leave all inter-molecular distances unchanged. In contrast, the remaining $N-6$ degrees of freedom are related to intramolecular vibrations (phonons, ``relative modes''). It is easy to verify that the Hamiltonian of this system of $N$ molecules can be split in a center of mass and a relative part which are independent (they commute). Therefore, only the vibrations contribute to the surface temperature $T^i_s$. On the other hand, the kinetic temperature (energy) is connected only to the center-of-mass (COM) modes -- in fact, only to the translational degrees of freedom. Now, in case of laser heating, predominantly the COM modes are excited via the radiation pressure, cf. the discussion in Sec.~\ref{ss:laser-matter}. This is due to the large spot size compared to the intermolecular distances (in order to excite the relative modes there should exist a substantial field gradient on the scale of the intermolecular separation which is not the case). Still, an open question is to what degree individual photons are absorbed by the molecules which could lead to an excitation of the relative degrees of freedom and, eventually, to a slight increase of the surface temperature, but this effect will be neglected here.

The surface temperature is almost the same for all particles (provided they are under the same plasma conditions), $T_s^i\approx T_s$, because it arises from a very large number of coupled modes of many particles $N$. In contrast,
the kinetic energies of individual particles (i.e. the individual $T_i$) are, in general, different. This is because the displacements of individual dust particles caused by lasers are, to a large extent, random, as is the case with the random displacements of the molecules of a gas. And as in the case of a gas, 
collisions between all dust grains eventually drive the system to thermodynamic equilibrium (with respect to the collective degrees of freedom). For a classical system one would expect that a Maxwellian velocity distribution is established the width of which then reflects the overall kinetic temperature $T$. This expectation is fully confirmed by our simulations and the laser heating experiments discussed below, cf. Fig.~\ref{fig:vspectra}.

The decoupling of center of mass and relative degrees of freedom and, correspondingly, of $T^i$ and $T^i_s$ (and, therefore also $T$ and $T_s$), has been verified--indirectly--experimentally. Here we mention detailed studies of the surface temperature by Maurer \textit{et al.}~\cite{maurer_heating_2011} for dust particles doped with a temperature sensitive fluorescent dye.
They concluded, from an energy balance model, that $T_s^i$ scales roughly linearly with the rf power.
At low rf discharge powers, $T_s^i$ has been found slightly above room temperature.
In contrast, in the experiments reported below, the kinetic temperature is found significantly higher than room temperature and thus it may strongly exceed $T^i_s$. 
This is because the radiation pressure from the laser(s) implies a substantial momentum transfer to the
particles. Combined with their small weight and low friction in the plasma, these laser beams
are capable to significantly accelerate individual dust particles. 

Returning to our original goal -- a control of $\Gamma$ -- we conclude that only the kinetic temperature has the meaning of thermodynamic temperature. (We mention that this holds rigorously only in the thermodynamic limit, however, many simulations and experiments  indicate very similar behavior in the case of finite clusters.) So, in the following we will concentrate entirely on the kinetic temperature $T$. We expect that variation of $T$ (and hence $\Gamma$) gives access to the thermodynamic properties of finite dust clusters and to phase transitions. The only requirement which we have to fulfill is that the employed heating scheme
should guarantee a truly random character of the individual kinetic temperatures $T^i$. In other words, the excitation should act like a thermostat for dust particles allowing for equilibration by particle scattering. As we will see below, properly chosen laser heating fulfills this requirement very well and allows for spatially homogeneous and stationary heating.


Finally we mention that, unlike $T_s$, the kinetic temperature $T$ cannot be obtained experimentally from a conventional heat flux analysis. Recent investigations by Fisher \textit{et al.}~\cite{fisher_thermal_2013} state that electrostatic fields in the plasma background provide a significant contribution to the kinetic temperature of dust particles. 
However, a complete understanding of the involved processes is still missing and this makes it difficult to control temperature this way. These questions are beyond the present review and are not crucial for our subsequent analysis.
In the experiments discussed below, the kinetic temperature (kinetic energy) of individual particles is recorded directly by tracking all particles. This gives access to the collective properties of the COM degrees of freedom of the entire dust cluster, including the power specturm and the velocity distribution.

\subsection{Realizing laser heating of dust particles}\label{ss:laserheating}
The idea to transfer momentum from a laser beam to a dust particle goes back to the early days
of dusty plasma research~\cite{homann1997, annaratone1997, takahashi1998, piel2002,  
homann1998, homann1999} and has been used for many purposes so far. This includes the investigation of
particle interaction potentials \cite{melzer2000, samarian2009, kroll2010}, the excitation of waves~\cite{homann1997,homann1998}, study of Mach
cones~\cite{melzer_laser-excited_2000, nosenko_observation_2002,nosenko2003, piel2006} or the
stability and normal mode analysis~\cite{klindworth_laser-excited_2000}. The first systematic laser
heating experiments were performed by Wolter \textit{et al.}~\cite{wolter_laser_2005} and by
Nosenko \textit{et al.}~\cite{nosenko_laser_2006, nosenko_heat_2008}.
In the experiment of Wolter \textit{et al.}~\cite{wolter_laser_2005} the spot of a laser beam is rapidly moved via scanning mirrors to one position in a 2D dust cluster and remains at this position for about one tenth of a second, accelerating dust grains during this time.
Then, the laser spot is rapidly moved to the next randomly chosen position via galvanometer mirrors. 
A similar heating scheme has been developed in Greifswald in order to manipulate finite 3D clusters \cite{Schella11}.
There, a laser beam is used to manipulate the cluster from the horizontal direction using this ``point and shoot'' technique.
In this experiment, a near Maxwellian velocity distribution of the particles is realized, but with different kinetic temperatures in the beam direction and perpendicular to it.

The experiment of Nosenko \textit{et al.}~\cite{nosenko_laser_2006, nosenko_heat_2008} exploits two opposing laser beams which were directed onto a 2D plasma crystal using scanning mirrors.
Driving these scanners with sinusoidal signals at an irrational frequency ratio a well defined area of the plasma crystal is scanned (i.e. heated) in a Lissajous figure-like fashion.
In addition, the opposing beam setup assures that the transfered momenta cancel on average, while the kinetic energy of each particle is raised as a result of non-compensated momentum fluctuations. Nosenko \textit{et al.} showed that their laser heating results in a Maxwellian velocity distribution parallel and perpendicular to the optical axis.
However, the temperature in perpendicular direction was found to be significantly lower.
Obviously, the viscous damping of the neutral gas impedes that collisions redistribute sufficient energy in perpendicular direction.

\begin{figure}
  \includegraphics[width=0.98\textwidth]{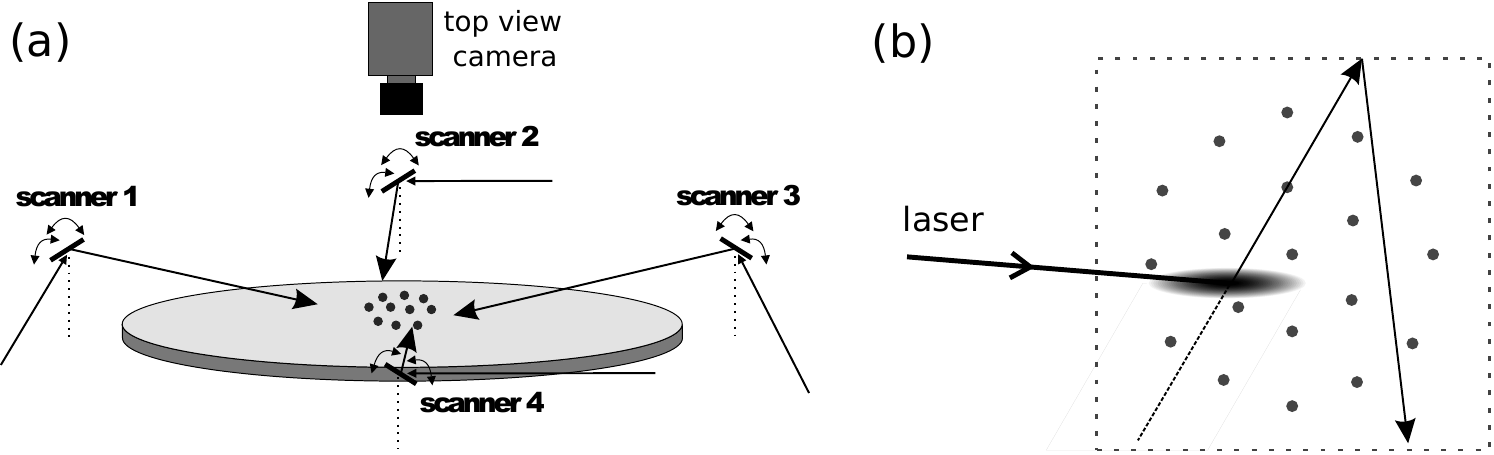}\\
  \caption{(a) Setup for a laser heating experiment using 4 laser beams.
  The spot of each laser beam is moved through the 2D plasma crystal layer
  by independently controlled scanning mirrors. The scanned area is large
  enough to cover the entire crystal. The radiation pressure force impels
  the particles in $\vec{e}_\parallel$ direction. (b) The improved heating
  method moves the spot with constant speed in both directions until the
  border of the scanned area is reached. At this point, a new speed is
  chosen randomly for the inverse direction.}
  \label{fig:setup_2d}
\end{figure}

To overcome this limitation recent laser heating experiments for 2D clusters use four laser
beams, where each optical axis is equipped with two opposing beams and perpendicular
orientation of both optical axes~\cite{schablinski_laser_2012,thomsen_laser_2012}, see
Fig.~\ref{fig:setup_2d}(a). In addition, the scanning procedure has been optimized. The reason
for this is that a scanning scheme based on Lissajous figures results in a velocity power
spectrum where strong harmonics of the scanning frequencies are observed. This is an
indication that the periodicity of the driver causes individual particles to move with the same
periodicity, i.e. between two kicks of the laser the particle velocity decreases significantly.
Therefore, the optimized scanning procedure [see Fig.~\ref{fig:setup_2d}(b)] assures that each
particle is driven by a laser beam before its motion originating from the previous laser drive is
damped out. Thus, the requirement that each spatial position is covered by the scanning
procedure is only sufficient if the maximum time between two complete scans is less than the
inverse of the damping rate. 

\begin{figure}
  \includegraphics[width=0.98\textwidth]{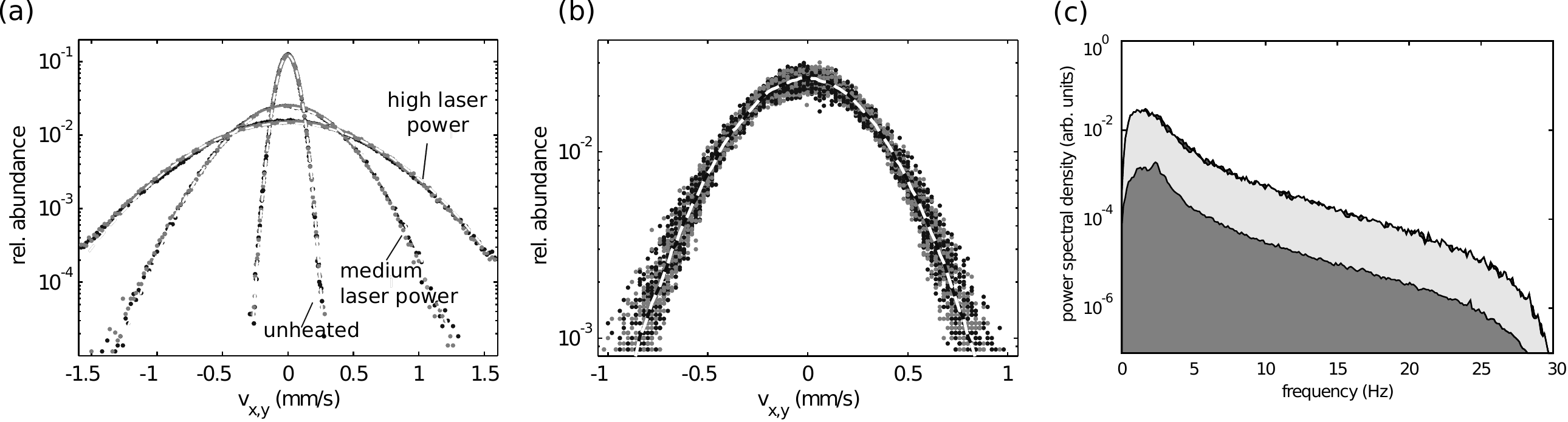}\\
  \caption{(a) Average particle velocity distribution function for a
2D cluster of 19 particles for three different values of the heating power. The velocity
components in $x$ and $y$  direction (gray and black dots)
are plotted separately. For each  case (unheated, medium and high
heating power) Gaussian fits (dashed lines) indicate only
slight deviation from an ideal Maxwell distribution. (b) Velocity
distributions for a heated system [similar to (a)], but here
the distribution is calculated for each particle independently.
The small scatter of the data confirms that all particles are heated equally.
(c) Power spectrum of the velocity fluctuations. The spectrum
of the unheated system is plotted in dark gray. The spectra of the
$x$ and $y$ component of a heated system are plotted (light gray and black).
Note that both components are equally heated and no artefacts from the
scanning process can be found in the spectra. From Ref.~\cite{schablinski_laser_2012}}
  \label{fig:vspectra}
\end{figure}

The experiments of Schablinski \textit{et al.} demonstrated that this
is feasible for small clusters ($N<100$). Their measured velocity spectra show no residual
peak of the scanning lasers. 
Fig.~\ref{fig:vspectra} summarizes the basic features of their
heating method. In Fig.~\ref{fig:vspectra}(a) the velocity distributions of the heated system are
clearly Maxwellian. Plot (b) stresses that the Maxwellian character is even obtained if the
velocity distribution is checked for each particle individually. The small scatter in particle
temperature is a clear indication that the heating process is spatially homogeneous. Therefore,
the laser heating with four laser beams and an optimized scanning procedure can be regarded
as an ideal thermostat for 2D dusty plasma crystals.

\section{Theoretical description of 2D laser heating}\label{s:numerics}
In this section, we first describe how the 2D laser manipulation experiment described above is simulated with the Langevin molecular dynamics (LMD) method~\cite{schablinski_laser_2012,thomsen_laser_2012,kudelis_heat_2013} and then develop an analytical model for the achieved temperature.
We explicitly study the elaborate heating scheme used in the experiments that will be presented below in Sec.~\ref{sec:2Dexp}.
The purpose of such computer simulations is to suggest optimal values for the heating parameters and to recommend future experiments like inhomogeneous heating.
Moreover, computer simulations provide the possibility to systematically scan single parameters like the heating power or the beam (spot) size at constant other parameters which is often experimentally too costly.

\subsection{Langevin Molecular Dynamics simulations}\label{ss:lmd}
In an LMD simulation, the Langevin equations of motion for the dust component
\begin{equation}
 m \frac{d\vec{v}_i}{dt} = \vec{F}_i + \vec{f}^{\rm L}_i - \gamma m \vec{v}_i + \vec{\xi}_i(t) {\rm ,} \quad i=1\dots N, 
 \label{eq:langevin}
\end{equation}
for all dust grains $i$ are integrated numerically. In these equations, the first term on 
 the right hand side results from minus the gradient with respect to $\vec{r}_i$ of the Hamiltonian, Eq.~(\ref{eq:dust_hamilton})
[i.e. the forces due to external potentials and inter-particle interactions that appear in Newton's equations of motion].  
The last two terms on the right hand side describe the frequent collision of the dust grains with
the neutral gas background statistically by a viscous damping force, $-\gamma m \vec{v}_i$, and
a random force $\vec{\xi}_i(t)$. The random force has zero average and can be modeled by a
Gaussian probability distribution with correlation function
\begin{equation}
 \left\langle \xi_{i,\alpha}(t) \xi_{j,\beta}(t')\right\rangle = 2 \gamma m \kB T \delta_{i,j} \delta_{\alpha,\beta} \delta(t-t') {\rm ,}
 \label{eq:random_force_correlation}
\end{equation}
where $i,j$ are particle indices and $\alpha,\beta \in\lbrace 1,2\rbrace$ indicate the spatial components of the random force vector. The amplitude of this force depends on the temperature $\kB T$ as well as on the friction coefficient $\gamma$ via the fluctuation-dissipation theorem (\ref{eq:random_force_correlation}).

The dust-dust interaction is described by a Yukawa potential, Eq.~(\ref{eq:Yukawa}).
In experiments on 2D clusters, the strong vertical confinement in the plasma sheath allows the formation of monolayer clusters.
Therefore, the system is treated strictly two-dimensional in the
simulation.

The effect of the heating lasers on each particle is given by the second term on the r.h.s. in Eq.~(\ref{eq:langevin}), $\vec{f}^L_i = \sum_{l=1}^{N_L}\vec{f}^L_{il}$, which is composed of separate contributions of all $N_L$ laser beams. 
The dominating effect of every single heating laser beam (index $l$) on particle ``i'' is a momentum transfer by the
radiation pressure which is described by the force $\vec{f}^L_{il}$.
If the laser spot hits a dust grain, this particle is accelerated in beam direction.
The laser force
\begin{equation}
 F_l = q_{\rm optic} {n_1} \frac{\pi r_{\rm p}^2 I_{\rm laser}}{c}
    = q_{\rm optic} {n_1} \frac{\pi r_{\rm p}^2}{c} \frac{P_{\rm laser}}{2\pi\sx\sy} \exp{\left[ -\frac{\Delta_x^2}{\sx^2}-\frac{\Delta_y^2}{\sy^2} \right]}
 \label{eq:laser_force_01}
\end{equation}
is proportional to laser intensity $I_{\rm laser}$ at the particle position $(\Delta_x,\Delta_y)$
relative to the laser spot center, the cross-section of the dust particle $\pi r_{\rm p}^2$,
the refractive index of the plasma $n_1$ surrounding the particle and a dimensionless quality factor $q_{\rm optic}$~\cite{liu_radiation_2003}.
This factor describing the momentum transfer by an incoming photon has the limiting values 1, if the particle was a perfect absorber, and  2 for a perfectly reflecting flat disk perpendicular to the beam.
For a spherical particle reflecting, absorbing and transmitting photons it is $q_{\rm optic}<2$.
Typical values of $q_{\rm optic}\approx 1$ are reported for melamine particles~\cite{liu_radiation_2003}.
$\sigma_{x,y}$ characterizes the extensions of the elliptical spot which has an area of $2\pi \sigma_x \sigma_y$.

As sketched in Fig.~\ref{fig:setup_2d}, the laser beams strike the cluster from above the levitation
plane with a low angle of incidence. The out-of-plane component of the accelerating force is
considered to have no impact on the particles' motion due to the strong vertical confinement.
However, the spot profile is stretched in beam direction due to the relatively small angle of incidence, $\alpha <
90^{\circ}$. The laser force is time and space dependent according to the experimentally chosen scheme. The amplitude of the
force depends on the particle's position inside the spot which is described by an anisotropic
Gaussian intensity profile. The force acting on a particle at position $\{\Delta_x(t),\Delta_y(t)\}=\vec{r}-\vec{r}_l(t)$ away form the moving spot center is described by
\begin{equation}
 \vec{f}_{l}\left(\vec{r},t\right) = \frac{P_{0}}{2\pi \sigma_x \sigma_y} \cdot \exp\left[ -\frac{\Delta_x^2(t)}{2\sigma_x^2} - \frac{\Delta_y^2(t)}{2\sigma_y^2} \right] \vec{e}_l \\ ,
 \label{eq:laserforce}
\end{equation}
where $\vec{e}_l$ is a unit vector in beam direction.
Here, based on Eq.~(\ref{eq:laser_force_01}), we have introduced dimensionless units for all lengths and $P_0$.
The amplitude of the force $P_{0}=q_{\rm optic} {n_1} {\pi r_{\rm p}^2 P_{\rm laser}}/{c}$ 
is determined by the laser power, the cross section of the dust
grain and its absorption and reflection characteristics. Since only the in-plane component has an
effect on the particle's motion, $P_0$ is reduced by the factor $\cos \alpha$ where $\alpha$
is the angle of incidence. The trajectories of the laser spot centers within the levitation plane are denoted by
$\vec{r}_l(t)=\left\{x_l(t),y_l(t)\right\}$ and depend on the heating scheme.

\subsection{Comparison of different heating schemes}\label{ss:lmd-schemes}
\begin{figure}
 \includegraphics{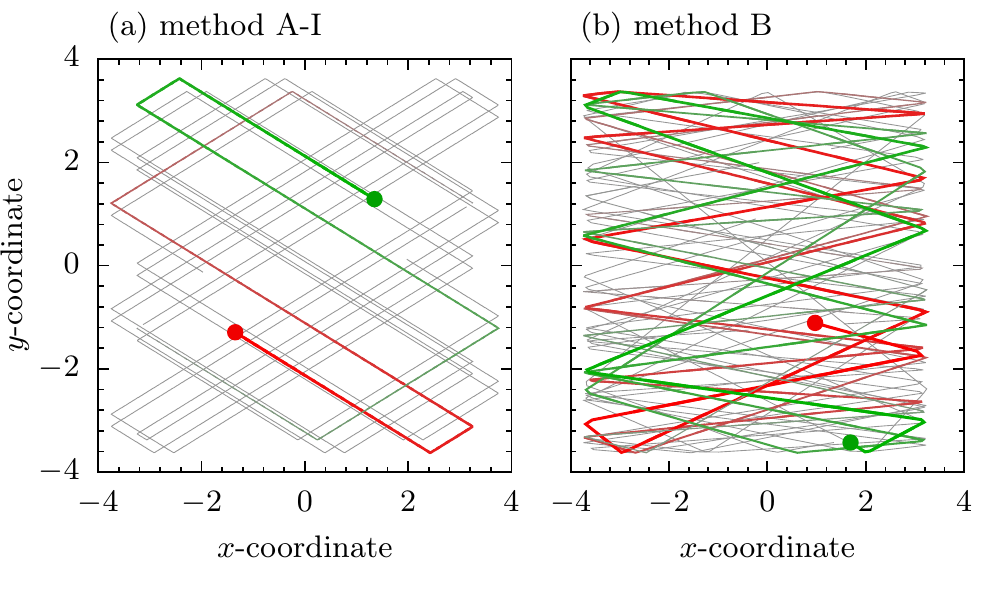}
 \caption{Trajectories of the laser spots (red/green: $\pm x$-direction)  for two of the investigated heating methods.
 (a) Method A-I uses a pseudo-irrational frequency ratio. (b) Using method B, a new scanning frequency is randomly 
 chosen each time a laser spot reaches the border. Here, the two additional spots of the laser oriented 
 in $\pm y$-direction are omitted, for the sake of clarity.\label{fig:LaserPattern}}
\end{figure}

%
All investigated heating methods (a more extensive analysis was reported in Refs.~\cite{thomsen_diplom,thomsen_laser_2012}) use triangular signals to drive the $x$- and
$y$-oscillations of the lasers spots $x(t)=x_0\cdot{\rm triag}(f_x t)$, $y(t)=y_0\cdot{\rm
triag}(f_y t)$. Using a sinusoidal signal would cause an increased intensity at the borders of the
scanned area. The heating methods differ in the scanning frequencies $f$. These frequencies are
fixed for heating methods~A-I. Method~A-I uses the same frequencies for both laser and a
pseudo-irrational ratio $f_x/f_y$. For method~B, the scanning frequencies are dynamically
changed each time a laser spot reaches the border of the scanned area.
Hence, no pattern is repeated~\cite{schablinski_laser_2012,thomsen_laser_2012}.
added.
The parameters of all heating methods are summarized in table~\ref{tab:LaserParameters} and the scanned patterns are shown in figure~\ref{fig:LaserPattern}.
\begin{table}
 \begin{tabular}{|l||c|c|}
    \hline
    method		& A-I			& B	\\ \hline
    laser beams		& 2			& 4	\\
    frequency ratio	& pseudo-irrational	& random frequencies	\\
    $f_\parallel$	& 14.5623\,Hz		& 50--300\,Hz	\\
    $f_\perp$		& 9.0\,Hz		& 15--60\,Hz\\
    \hline
 \end{tabular}
 \caption{Parameters defining the pattern which is scanned by the lasers. Since both beams are oriented in $\pm x$-direction for methods A-I, $f_\parallel=f_x$ in that case.\label{tab:LaserParameters}}
\end{table}

For the simulation results presented in this review, we used the following dimensionless units 
of length, time and energy: $r_0=\left[{Q_{\rm d}^2}/{(4\pi \varepsilon_0\, m\omega_0^2)}\right]^{1/3}$,
$t_0=\omega_0^{-1}$, and $E_0=\left[{Q_{\rm d}^4\,m\omega_0^2}/{(4\pi\varepsilon_0)^2} \right]^{1/3}$.
For the numerical integration of the stochastic differential equations, we used an integration scheme described 
by Mannella \textit{et al.} that can be considered as an extension of the ``leap frog'' scheme~\cite{mannella_quasisymplectic_2004}.
For details concerning the simulations, we refer the reader to Refs.~\cite{thomsen_laser_2012,kudelis_heat_2013}. 
(This scheme can also be easily generalized to incorporate homogeneous magnetic fields
of arbitrary strength~\cite{spreiter_classical_1999,chin_symplectic_2008,ott_wave_2013}).

As a first result, our LMD simulations confirmed the importance of randomly changing the scanner frequency. 
Especially rational scanning frequency ratios $f_x/f_y$ which result in a closed scanning pattern prove to be problematic.
%
A second result is the confirmation of the excellent heating qualities (with respect to homogeneity of the heating) of the random frequency method~B~\cite{thomsen_laser_2012}.
While constant scanning frequencies and combinations of them appear as peaks
in the power spectral density (PSD) of the particle velocities for the method~A-I, this artefact is removed by
randomly changing the scanning frequencies, see Fig.~\ref{fig:MDspectrum} and
Fig.~\ref{fig:vspectra}(c). The PSD further discloses that effective energy transfer from $x$- to
$y$-motion works for low frequencies, $f \lesssim 8\, {\rm Hz}$, only. The pseudo-irrational
frequency method~A-I shifts the entire spectrum of the motion in $x$-direction to higher
energies but also induces several peaks at the scanner frequencies and combinations thereof,
Fig.~\ref{fig:MDspectrum}(a). 

Both experiment, Fig.~\ref{fig:vspectra}(c), and LMD simulation,
Fig.~\ref{fig:MDspectrum} (bottom), show that the random frequency method B is well suited as
a thermostat for the 2D dust system. The energy input is homogeneous over the entire frequency
spectrum. It is also homogeneous in space and the Maxwellian shape of the velocity profile is
conserved in both $x$- and $y$-direction.
Moreover, the random frequency method~B is very robust against changes in the laser parameters~\cite{thomsen_laser_2012}.
The temperature increase $\Delta T$ (i.e. the heating effect) is reduced when the typical scanning frequency $\bar{f}$ of the laser beams is increased according to  $\Delta T \propto \bar{f}^{-1}$. 
However, in order to achieve a thermal heating effect, each particle should frequently be affected by a laser spot.
This condition is violated for $\bar{f}$ below the trap frequency $\omega_0$.
Therefore, scanning frequencies of the order of a few $\omega_0$ are advantageous~\cite{thomsen_laser_2012}.

\subsection{Analytical estimate for the temperature increase}\label{ss:analytics}
%
We now derive an analytical approximation for the heating power as a function of the laser parameters: 
the force amplitude $P_{0}$, spot size $\sigma_{x,y}$, size of the scanned array $2X\times 2Y$ and 
average spot velocity. The spot velocity is connected to the scanning frequencies, $f_{l,x}, f_{l,y}$, via $v_{l,x}=4X f_{l,x}$ and $v_{l,y}=4Y f_{l,y}$.
A similar analytical derivation of the heating effect was performed by Wolter \textit{et al.} for a setup with one laser spot that was rapidly moved to one position, remained there for a dwell time of a few microseconds and was then moved to the next position~\cite{wolter_laser_2005}. 

Here we extend this model to a continuously moving laser spot corresponding as closely as possible to the experimental heating method by Schablinski \textit{et al.} that is discussed in Sec.~\ref{s:exp_2d}.
For simplicity, the momentum transfer is computed in one direction only. 
Since the velocity of the dust 
is small compared to the velocity of the moving laser spot, the particle's displacement 
during an acceleration event can be neglected. The time-dependent force acting on a particle 
when a spot passes with distance $\Dyzero$ in $y$-direction is
\begin{equation}
 F(t') = \pm P_0 \frac{1}{\sqrt{2\pi}} 
    e^{-\frac{{t'}^2}{2\tau^2} 
      } \cdot 
    \underbrace{
      \frac{1}{\sqrt{2\pi}} \frac{1}{\sx \sy}
      \exp \left\lbrace
	-\frac{1}{2} \left[
	  \frac{\Dyzero^2}{\sy^2} - \tau^2 \frac{\Dyzero^2 v_y^2}{\sy^4}
	\right]
      \right\rbrace
     }_{:= \Sigma} {\rm ,}
\label{eq:force_time_06} 
\end{equation}
where we introduced the time scale of the passing event
$ \frac{1}{\tau^2} = \frac{v_{l,x}^2}{\sx^2} + \frac{v_{l,y}^2}{\sy^2} {\rm ,}$
and the shifted time $t'=t+\tau^2 \frac{\Dyzero v_{l,y}}{\sy^2}-t_0$. Here, $t_0$ is the time 
when the spot has the smallest distance $\Dyzero$ in $y$-direction and we introduced the time-independent 
geometry factor $\Sigma$. 
The total momentum transfer to the particle during this laser hit follows from Eq.~(\ref{eq:force_time_06})
\begin{eqnarray}
  \Delta p &=& \int_{-\infty}^{\infty}{ dt'\ F(t') }
      = \pm P_0 \Sigma \cdot \tau {\rm ,}
\label{eq:momentum_transfer_01}
\end{eqnarray}
and is canceled, on average, by the momentum transfer from the opposing laser beam. To calculate the average energy transfer, we need to average the squared geometry factor $\Sigma^2$ over all possible passing distances $\Dyzero$. 
Assuming that the particle at $(x_{\rm p},y_{\rm p})$ does not come too close to the boundary, the integration limits can be extended to infinity, with the result
\begin{eqnarray}
 \left\langle \Sigma^2 \right\rangle_y  
  &\approx& \frac{1}{2Y} \cdot \frac{1}{2\sqrt{\pi}} \cdot \frac{1}{\sx\sy} \cdot \frac{1}{v_{l,x}} \cdot \frac{1}{\tau}
  \label{eq:Sigma2_yav_05}  {\rm .}
\end{eqnarray}
Each time the laser crosses the levitation plane in $x$ direction, it passes the particle position once. 
Then, the ``passing rate'' $\chi$  is given by the inverse crossing time in $x$ direction as
$ \chi = 
\frac{v_{l,x}}{2X} {\rm .}$

The energy change of a particle with velocity $v$ during a single laser-particle interaction event is 
(neglecting the deceleration due to friction during the kick), 
\begin{equation}
  \Delta E = E' - E = v \Delta p + \frac{\Delta p^2}{2m} {\rm ,}
  \label{eq:energy_transfer_01}
\end{equation}
where the momentum transfer $\Delta p$ is given by Eq.~(\ref{eq:momentum_transfer_01}). 
From this the average energy transfer is obtained using Eq.~(\ref{eq:Sigma2_yav_05}),
\begin{equation}
\langle \Delta E \rangle = \frac{1}{2m} P_0^2 \Sigma^2 \cdot \tau^2 {\rm ,}
  \label{eq:energy_transfer_02}
\end{equation}
which depends on the passing distance $\Dyzero$ via $\Sigma$.
Then, the average energy transfer per time follows from averaging over $\Dyzero$, multiplying by $\chi$ and 
using Eq.~(\ref{eq:Sigma2_yav_05}),
\begin{eqnarray}
  \left\langle \frac{\Delta E}{\Delta t} \right\rangle &=&
  \frac{1}{2m} \cdot \frac{1}{2\sqrt{\pi}} \cdot  P_0^2 \cdot \frac{1}{4XY} \cdot \frac{1}{\sx \sy} \cdot \tau_{\rm eff} {\rm ,}
  \label{eq:energy_transfer_05}
\end{eqnarray}
where we also took into account that the laser spot velocity components $v_x$, $v_y$ vary in the experiment
by computing an effective time scale $\tau_{\rm eff}$ by averaging $\tau(v_{l,x},v_{l,y})=\left({v_{l,x}^2}/{\sx^2} + {v_{l,y}^2}/{\sy^2}\right)^{-1/2}$
over all possible spot velocities $v_{l,x} \in [v_x^{\rm min},v_x^{\rm max}]$ and 
$v_{l,y} \in [v_y^{\rm min},v_y^{\rm max}]$~\cite{tau-average}.

From the result~(\ref{eq:energy_transfer_05}), 
we can now calculate the equilibrium temperature $T_{\rm eq}$ in the Langevin MD model
which follows from the balance between laser input power $P_{\rm laser}$, the
power input by the stochastic force, $P_{\rm stochastic}$, and 
the power loss due to friction, $P_{\rm friction}$,
\begin{equation}
   P_{\rm laser} + P_{\rm stochastic} + P_{\rm friction} = 0 {\rm .}
  \label{eq:power_balance}
\end{equation}
Assuming a 1D Maxwellian particle velocity distribution with temperature $T_{\rm eq}$,
\begin{equation}
  p(v) = \sqrt{\frac{m}{2\pi \kB T_{\rm eq}}} e^{-\frac{m v^2}{2\kB T_{\rm eq}}} {\rm ,}
  \label{eq:Maxwellian_01}
\end{equation}
and the velocity change due to friction, $\dot{v}=-\gamma v$, where $\gamma$ is the friction coefficient, we obtain
\begin{eqnarray}
 P_{\rm friction} &=& \left\langle \frac{d}{dt} \frac{m}{2} v^2 \right\rangle
	= - m\gamma \left\langle v^2 \right\rangle
	= - \gamma \kB T_{\rm eq}
  \label{eq:power_loss_02}
\end{eqnarray}
On the other hand, in an equilibrium system without laser heating at the temperature $T_0$ of the neutral gas, the power loss due to friction is be compensated
by $P_{\rm stochastic}=\gamma \kB T_{0}$.
Using Eq.~(\ref{eq:power_balance}) and the power input by the laser from Eq.~(\ref{eq:energy_transfer_05}) multiplied by $2$ to take into account 
the pair of lasers in each direction, we obtain our final result for the equilibrium kinetic temperature of the dust particles
\begin{eqnarray}
  T_{\rm eq} &=& T_{0}
    +  \frac{\sqrt{\pi^3}}{2m\gamma \kB } \cdot I_0^2 \cdot \frac{\sx \sy}{XY} \cdot \tau_{\rm eff}
    \label{eq:power_balance_02} {\rm ,} 
\end{eqnarray}
where we introduced the laser intensity according to $I_0=P_0/(2\pi\ \sx\sy)$.
It is interesting to note the dependence on the relevant system parameters: $T_{\rm eq}$ grows proportional to the square of the laser intensity, in agreement with the findings of Wolter \textit{et al.}~\cite{wolter_laser_2005}, and the ratio of spot area to scan area and inverse proportional to the neutral gas friction.
These qualitative trends are also seen in the experiments.
The time scale $\tau_{\rm eff}$ introduces additional slightly more complex dependencies \cite{tau_note}.


\begin{table}
\newcolumntype{A}{>{\cellcolor[rgb]{0.9,0.9,0.9}}c}
\begin{tabular}{Ar|rrr}
  $\gamma$	&formula		& simulation			&deviation			&\\
  \toprule	
		&\multicolumn{3}{>{\columncolor[gray]{0.8}}c}{$\sx={0.50}$, $\sy={0.10}$}	& \\
  \midrule
		&${53.23}$		&${63.17}$		&				&$\Gamma$	\\
   \multirow{-2}{*}{${0.67}$}	
		&$18.79\times10^{-3}$	&$15.83\times10^{-3}$	&$15.7\% $			&$T$		\\
  \midrule
		&${70.24}$		&${80.26}$		&				&$\Gamma$	\\
  \multirow{-2}{*}{${1.00}$}	
		&$14.24\times10^{-3}$	&$12.46\times10^{-3}$	&$12.5\%$			&$T$		\\
  \midrule
		&${83.71}$		&${94.52}$		&				&$\Gamma$	\\
  \multirow{-2}{*}{${1.33}$}	
		&$11.95\times10^{-3}$	&$10.58\times10^{-3}$	&$11.4\%$			&$T$		\\
  \midrule
		&${94.65}$		&${104.7}$		&				&$\Gamma$	\\
  \multirow{-2}{*}{${1.66}$}	
		&$10.56\times10^{-3}$	&$9.550\times10^{-3}$	&$9.60\%$			&$T$		\\
  \midrule
		&${104.0}$		&${114.5}$		&				&$\Gamma$	\\
  \multirow{-2}{*}{${2.00}$}	
		&$9.619\times10^{-3}$	&$8.732\times10^{-3}$	&$9.22\%$			&$T$		\\
  \midrule
          	&${111.6}$		&${121.5}$		&				&$\Gamma$	\\
  \multirow{-2}{*}{${2.33}$}	
		&$8.965\times10^{-3}$	&$8.234\times10^{-3}$	&$8.15\%$			&$T$		\\
  \midrule
		&${118.0}$		&${127.7}$		&				&$\Gamma$	\\
  \multirow{-2}{*}{${2.66}$}	
		&$8.473\times10^{-3}$	&$7.831\times10^{-3}$	&$7.57\%$			&$T$		\\
  \bottomrule
 \end{tabular}
 \caption{Comparison of Eq.~(\ref{eq:power_balance_02}) and results from a LMD simulation.
 For smaller spots (not listed), the agreement is even better.\\
  Parameters: particle number $N=25$, 
  screening parameter $\kappa=1$,
  coupling parameter without laser heating $\Gamma_0=200$, 
  trap frequency $\omega_0=5.5\, {\rm Hz}$,
  dimensionless laser power $F_0=10$,
  scanned area $X=Y=3.5$
  \label{tab:power_compare}}
\end{table}
 
Despite the simplicity of the  model, the accuracy of the result (\ref{eq:power_balance_02}) is very good, 
as is confirmed by comparison to LMD simulations, see table~\ref{tab:power_compare}. 
The temperature is overestimated slightly, and the accuracy improves with $\gamma$ from about $15\%$ to better 
than $10\%$. For very small spot sizes and low friction (not shown) the agreement is even better 
than $5\%$.
For very large spot sizes, $\sigma_x \gtrsim 0.7$ in dimensionless units, the formula becomes
less accurate. Then, most particles are close to the border of the scanned area compared to 
the spot width, violating the assumption of Eq.~(\ref{eq:Sigma2_yav_05}). Moreover,
$\tau_{\rm eff}$ becomes large in this case and the deceleration due to friction as well
as the interaction with other particles during the hitting event should not be neglected.

Our result is not only in very good agreement with the simulations, it is also useful for characterizing 
the heating in the experiment. In fact, the spot size used in Tab.~\ref{tab:power_compare} is typical for the experiments.
Assuming that the visible spot width is $\Delta_{\rm vis}=3\sigma$, our
formula suggests a coupling strength of $\Gamma^{\rm form}=80$ which is in good
agreement with the measured temperature $T=3.5\, {\rm eV}$ corresponding to
$\Gamma^{\rm exp}=63$ for the maximum laser power used in Ref.~\textit{et al.}~\cite{schablinski_laser_2012}. However, that result has to be
understood as a rough approximation, since parameters like the spot size, the 
trap frequency, and laser power losses in the optical setup each have an uncertainty of
several percent in the experiment.

\begin{figure}
 \includegraphics{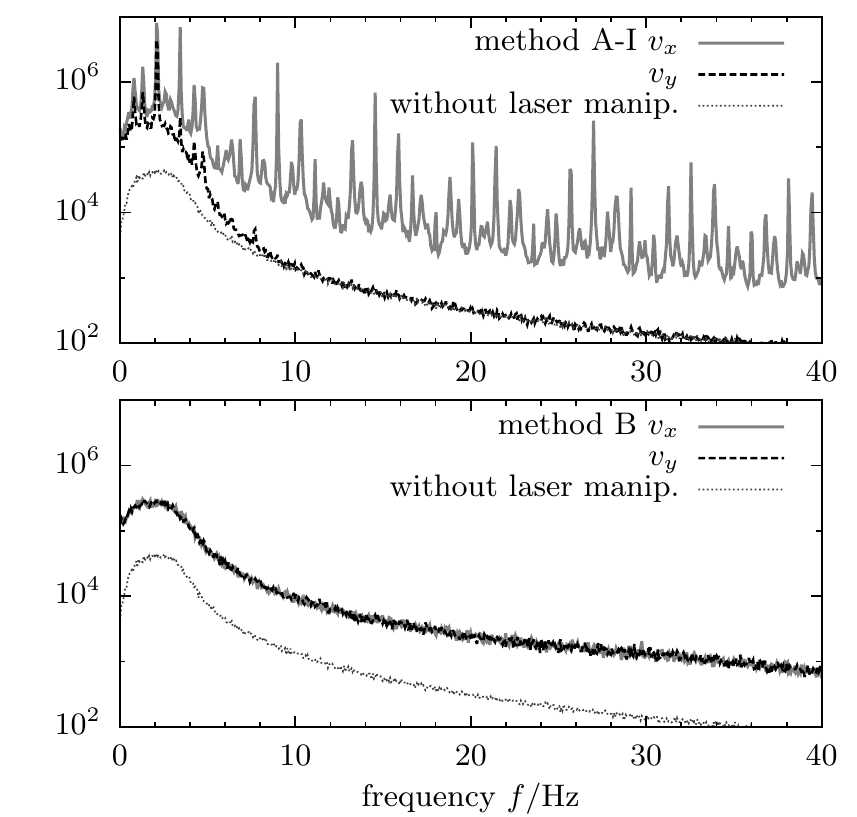}
 \caption{\label{fig:MDspectrum}Power spectral density (PSD) of the dust velocity averaged 
 over all particles of a Yukawa cluster with $N=38$ particles with different heating methods 
 (trap frequency $\omega_0=5.5\,{\rm s}^{-1}$). \textbf{Top:} Method A-I uses one pair of 
 laser beams in $\pm x$-direction and a pseudo-irrational scanning frequency ratio $f_x/f_y$.
 \textbf{Bottom:} Heating method B uses two pairs of laser beams to accelerate the particles 
 in both $\pm x$- and $\pm y$-direction. The scanning frequencies are randomly chosen each 
 time the border of the scanned area is reached. The spectra for $v_x$ and $v_y$ coincide,
 here. Only this method conserves the shape of the PSD by shifting the entire spectrum
 to higher energies.}
\end{figure}

\section{Experimental results for two-dimensional dusty plasmas}\label{s:exp_2d}
We now turn to the experiments on laser heating. First, we start with two-dimensional systems and
continue, in Sec.~\ref{s:3d}, with experiments with finite three-dimensional dust cluster.
Before presenting our experimental results for finite 2D dust clusters we briefly summarize
previous work on extended systems.
\subsection{Experimental results for extended systems}\label{sec:2Dexp_ext}
In many experiments in extended 2D systems the Lissajous heating scheme A-I (cf. Sec.~\ref{ss:lmd-schemes}) has been
applied since it allows one to densely scan (and heat) a well defined dust area. This technique has been
used successfully to drive 2D extended dust systems into the liquid state
\cite{nosenko_heat_2008,Nosenko09a,Feng08}. There, it has been demonstrated that the defect
concentration in steady-state laser-heating experiments exhibits an Arrhenius-type dependence
on the kinetic temperature \cite{Nosenko09a}. Moreover, these experiment suggest a
grain-boundary-induced melting scenario which is also observed in non-equilibrium heating
experiments, e.g. by changing the gas pressure of the plasma discharge \cite{melzer_experimental_1996}.

Further, experiments on heat transport and particle transport properties, such as diffusion
and viscosity, of laser-heated dust layers have been performed, see e.g.
\cite{Chan04,Nunomura05,Liu06,nosenko_heat_2008,Feng10,feng_observation_2012}. From these, fundamental transport
parameters like the thermal diffusivity of the dust component, as well as diffusion constants,
$\Gamma$-dependent viscosities and anomalous diffusion properties have been identified and
measured. These experiments have yielded important information 
on the application of laser-dust interaction methods in the field of dusty plasmas 
which are of high relevance also for strongly correlated small dust clusters at finite temperature.

\subsection{Experimental results for 2D clusters\label{sec:2Dexp}}

\begin{figure}
  \includegraphics[width=0.95\textwidth]{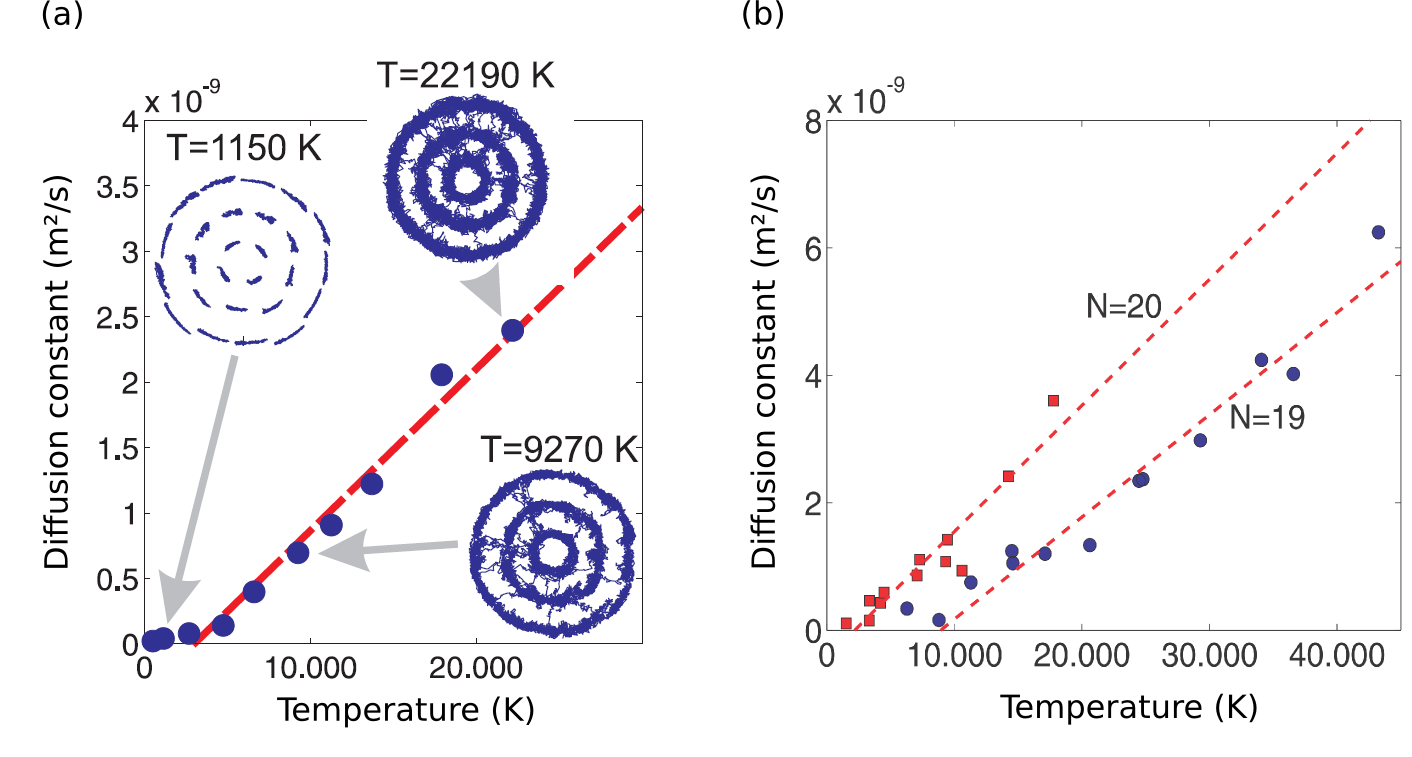}
  \caption{Diffusion constant as a function of temperature for finite 2D clusters. (a) Above a critical temperature
  the diffusion constant increases linearly with temperature. The insets show trajectories of a dust cluster with $N=26$
  particles at different temperatures. (b) Diffusion constant for two different dust clusters.
  The $N=19$ cluster with its (1-6-12) shell occupation is highly symmetric, and its critical temperature is significantly
  higher than that of the less symmetric $N=20$ cluster. (From Ref. \cite{Melzer13}).
  \label{fig:diffusion}}
\end{figure}

The reminder of this
section will concentrate on finite 2D clusters and their thermodynamic properties. Especially the
phase transition of these small systems is of interest as it should significantly depend on the
cluster size. The following experiments use the isotropic heating method B of Schablinski
\textit{et al.}~\cite{schablinski_laser_2012,thomsen_laser_2012}, which has been introduced in
Sec.~\ref{ss:lmd-schemes}. As shown there, the laser heating effectively provides a heat bath for the dust particles 
assuring that the dust subsystem is in thermodynamic equilibrium. Changing the laser power will result in a different
temperature of this thermostat. In the experiment, for each temperature, long time series are recorded to obtain the
trajectories of all particles. Examples of such particle trajectories are shown for low, medium
and high temperature in Fig.~\ref{fig:diffusion}(a). For low temperature the particles are well
localized and only a slight angular rotation is observed. At medium temperature the particles are
less localized in angular direction. Finally,  at high temperatures (high laser power), the radial correlations vanish as well.
Thus, we confirm that the melting process has two phases: first, a loss of angular correlation and, second, a loss
of radial correlation~\cite{Bedanov94,Melzer12}.

To determine the melting temperature several methods have been proposed (see
\cite{boning_melting_2008} and references therein). Unfortunately most of them either fail for
small clusters or are experimentally not feasible since they require extremely long time series in order to achieve sufficient
statistics, see Sec.~\ref{ss:tdyn}. Therefore, recently different methods have been applied which
were introduced in Secs.~\ref{ss:quantities}, \ref{ss:transport}: The first is the Instantaneous
Normal Mode (INM) analysis. This method computes the frequencies of the eigenmodes of a
cluster from the eigenvalues of the dynamical matrix, see Sec.~\ref{ss:transport}.
The results of such an INM-analysis are plotted in Fig.~\ref{fig:diffusion}.
The plots show that, above a critical temperature, the diffusion constant increases linearly with temperature.
A freezing temperature can be derived approximately from the point where $D$ vanishes
\cite{Das04}. 
Thus one can estimate a melting temperature $T_M$ by extrapolating the $D(T)$ curve toward zero.
Especially in Fig.~\ref{fig:diffusion}(b) the comparison of a highly symmetric cluster ($N=19$) and a cluster with low symmetry ($N=20$)
reveals that the melting temperature of the symmetric cluster is significantly higher ($T_{\rm M}^{19} \approx 9.000\,{\rm K}$,
$T_{\rm M}^{20} \approx 2.000\,{\rm K}$). A systematic investigation of melting temperatures as a function of
particle number \cite{Melzer13} has shown that symmetry has a mayor influence on melting temperatures of finite
systems, confirming earlier theoretical predictions \cite{Bedanov94,filinov_prl01}.


\begin{figure}
 \includegraphics{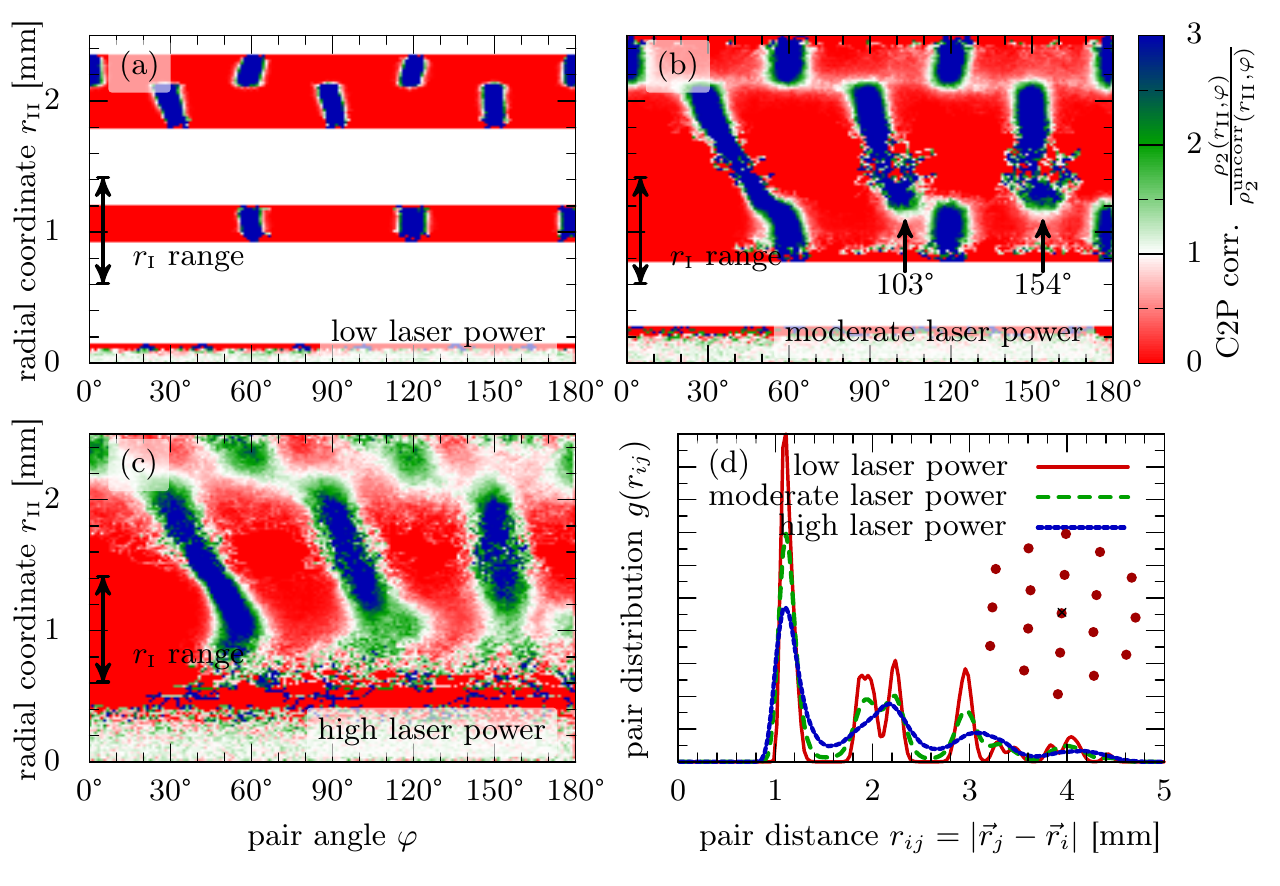}
 \caption{\label{fig:c2p_exp} (a)-(c) C2P correlation function for a laser heated cluster of $N=19$ particles. 
 The low laser power corresponds to $T$=2.800\,K, the moderate (high) power to $T$=17.000\,K ($T$=34.000\,K). 
 The first radial coordinate is averaged over the inner shell (arrow).
 (d) Radial pair distribution function for all three laser powers.
 Note that it does not distinguish between different shells. 
 The inset shows the cluster configuration at the lowest laser power.}
\end{figure}
The above results show that the INM analysis is sensitive to the cluster symmetry. However, it
does not resolve different melting processes such as inter-shell and intra-shell melting. 
(For example, it is well known \cite{Bedanov94} that the above two clusters have angular
melting temperatures that differ by many orders of magnitude but that their radial
melting temperatures are comparable.) 
For this reason, we also consider the center-two-particle correlation [cf. Sec. \ref{ss:quantities}] for the
experimental cluster of 19 particles. Figure~\ref{fig:c2p_exp}(a) shows a highly ordered structure of the
weakly manipulated cluster. The inner shell is occupied by six particles. Intra-shell neighbors
are found under angles of  $\gangle=60^\circ$, 120$^\circ$ and 180$^\circ$, as is
clearly seen by the peaks at these angles and a radius $\rtwo\approx 1{\rm mm}$, corresponding to the inner
shell. The outer shell is occupied by twelve particles and the angular order with respect to the
inner shell is fixed. Distinct peaks are visible at multiples of $\gangle=30^\circ$. When the
laser power is increased to a moderate value (cf. Fig.~\ref{fig:c2p_exp}(b)), clear intra-shell and
inter-shell correlations persist. However, inter-shell rotation (``angular melting'') has started, as
the peaks at the outer radius are no longer fully separated. At the inner shell radius, weak peaks
at $\gangle\approx103^\circ$ and $\approx 154^\circ$ (arrows in Fig.~\ref{fig:c2p_exp}(b))
indicate the occurrence of configurations with seven particles on the inner shell. The
appearance of this metastable configuration indicates the onset of radial melting. Finally,  at
high laser power (Fig.~\ref{fig:c2p_exp}(c)), the correlations between inner and outer shell have
vanished almost completely. At this heating power, the frequent particle transitions between the
two shells give rise to a finite density in the region between these shells. 

Thus the C2P correlation
function fully confirmed the stability of this ``magic number'' cluster against inter-shell rotation.
Yet a complete quantitative analysis of the different melting temperatures and their dependence on the
particle number is still open.

\section{Experimental results for 3D clusters}\label{s:3d}
\begin{figure}
  \includegraphics[width=107mm]{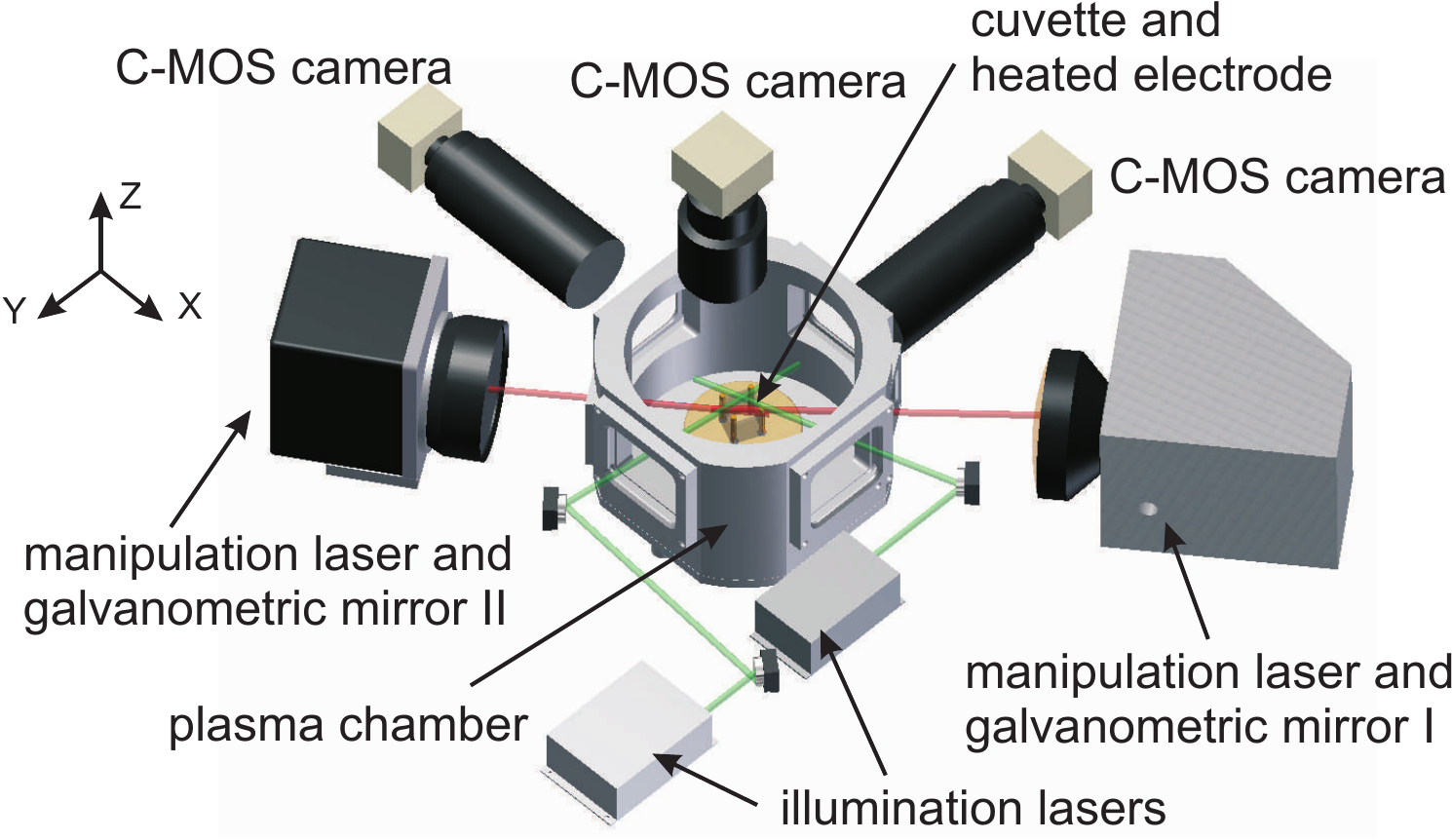}\\
  \caption{Sketch of the experimental setup. The particles are trapped in
the cubic glass box and are illuminated by two Nd:YAG lasers from two sides and
heated by two diode lasers from opposite directions. The three orthogonal
high-speed cameras allow to trace the full 3D particle motion individually. (From Ref.~\cite{Schella13a}.)
\label{fig:setup_3d}}
\end{figure}

Three-dimensional dust clusters are formed in parallel plate radio-frequency (rf, 13.56\,MHz) discharges, see Fig.~\ref{fig:setup_3d} and e.g. Ref.~\cite{Arp04, Kaeding08, annaratone_complex-plasma_2004, antonova_study_2009}.
The discharges are typically operated in argon at gas pressures between 1 and 100\,Pa and at rf powers between 1 and 10~W. 
The dust grains trapped in the discharge generally are monodisperse plastic microspheres with diameters chosen between 3 and 10\,$\mu$m.
These three-dimensional clusters are trapped inside a cubic glass cuvette placed onto the lower
electrode. The glass box provides inward electric forces on the negative dust grains. To
compensate the gravitational force an upward thermophoretic force is applied by heating the
lower electrode. The combination of all forces provides a 3D harmonic confinement
\cite{Arp05}. There, the particles arrange in nested spherical shells forming Yukawa balls or
Yukawa clusters \cite{Arp04,Block08,Kaeding08}, cf. Sec.~\ref{ss:structure}.

In the experiments described here, the particles are illuminated by low-intensity laser beams
and the scattered light is recorded with high-speed video cameras at frame rates of 50 to 200
frames per second (fps), typically \cite{Block08,Kaeding08,Schella11,Schella13}. For the
observation of the 3D clusters a stereoscopic camera setup \cite{Kaeding08} is used where the
particles are observed from three orthogonal directions. This setup allows to measure and
reconstruct the full 3D trajectories of clusters with up to $N=100$ particles with high temporal
resolution~\cite{Kaeding08,Himpel11}. Consequently, the dynamical properties of the dust
cluster can be followed for all particles individually.
\begin{figure}
  \includegraphics[width=107mm]{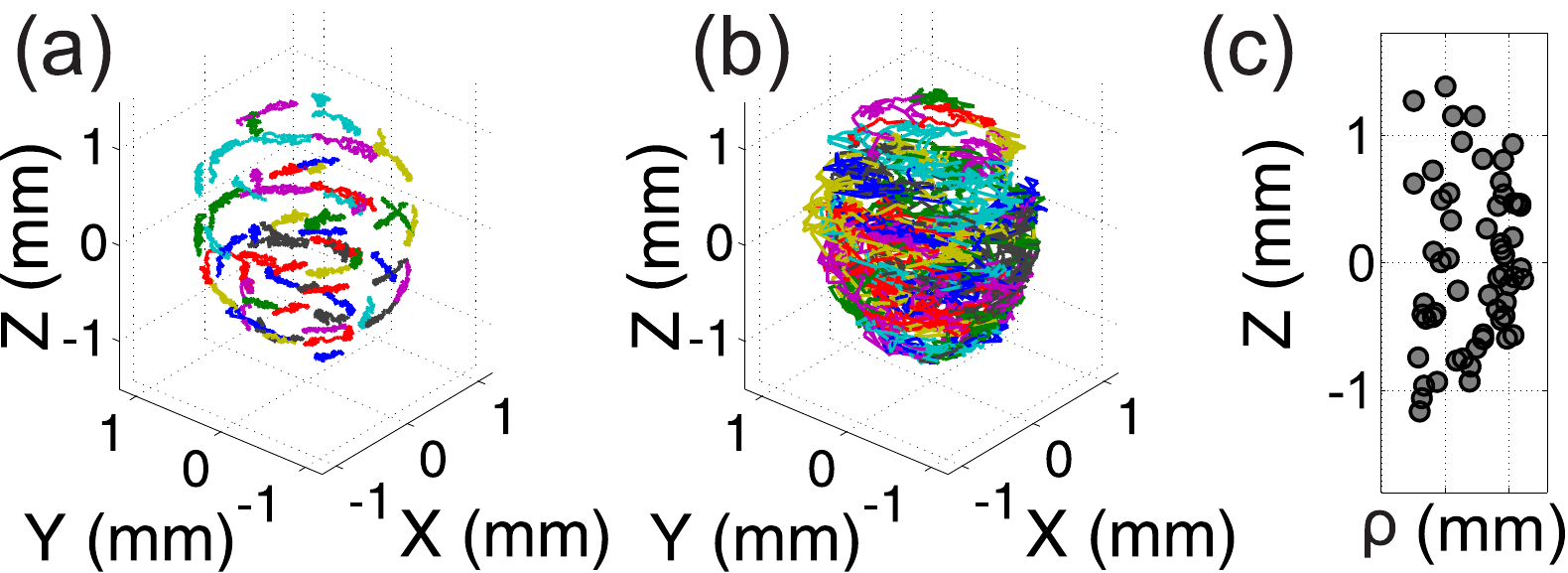}\\
  \caption{Trajectories of a $N = 60$ particle
cluster recorded over a time span of about 10\,s. (a) Without laser excitation, (b) with 250\,mW laser power. (c) Equilibrium
particle positions in cylindrical coordinates $\rho = \sqrt{x^2+y^2}$ and $z$
for the unheated case (a). (From Ref.~\cite{Schella13a}.)}\label{fig:traj}
\end{figure}
To realize a heating process for these 3D clusters a simpler approach than for the 2D case had to
be chosen due to experimental constrains.
The transfer of the elaborate heating schemes used for the 2D clusters (Sec.~\ref{sec:2Dexp_ext}) to the 3D case would require six beams, two of which would be blocked by the electrodes. 
Here, only two additional opposing laser beams are oriented parallel to the electrode and are randomly swept over the cross section of the cluster with a dwell time of $\tau = 0.1$\,s at each position~\cite{Schella11,Schella13,Schella13a}. 
The two laser beams at 660\,nm wavelength are operated with up to 1\,W output power. 
The random ``kicks'' to the particles by the radiation pressure mimic a heating process for these 3D clusters.

As a consequence, the resulting velocity distribution of the particles in the cluster is only
near-Maxwellian, with an overpopulation of ``cold'' dust particles \cite{Schella11}. Also the
heating is more effective in the direction of the beams resulting in higher temperatures along
this axis. Consequently, this laser-heating scenario of 3D clusters does not provide a true
thermodynamic heating, yet. Nevertheless, from the velocity distributions reasonable kinetic
temperatures $kT_{\alpha} = m\langle v_\alpha^2\rangle$ with $\alpha={x,y,z}$ can be assigned
and values of the order of a few times $10^4$\,K (few eV) have been realized by this laser
heating setup.
%

\begin{figure}
  \includegraphics[width=127mm]{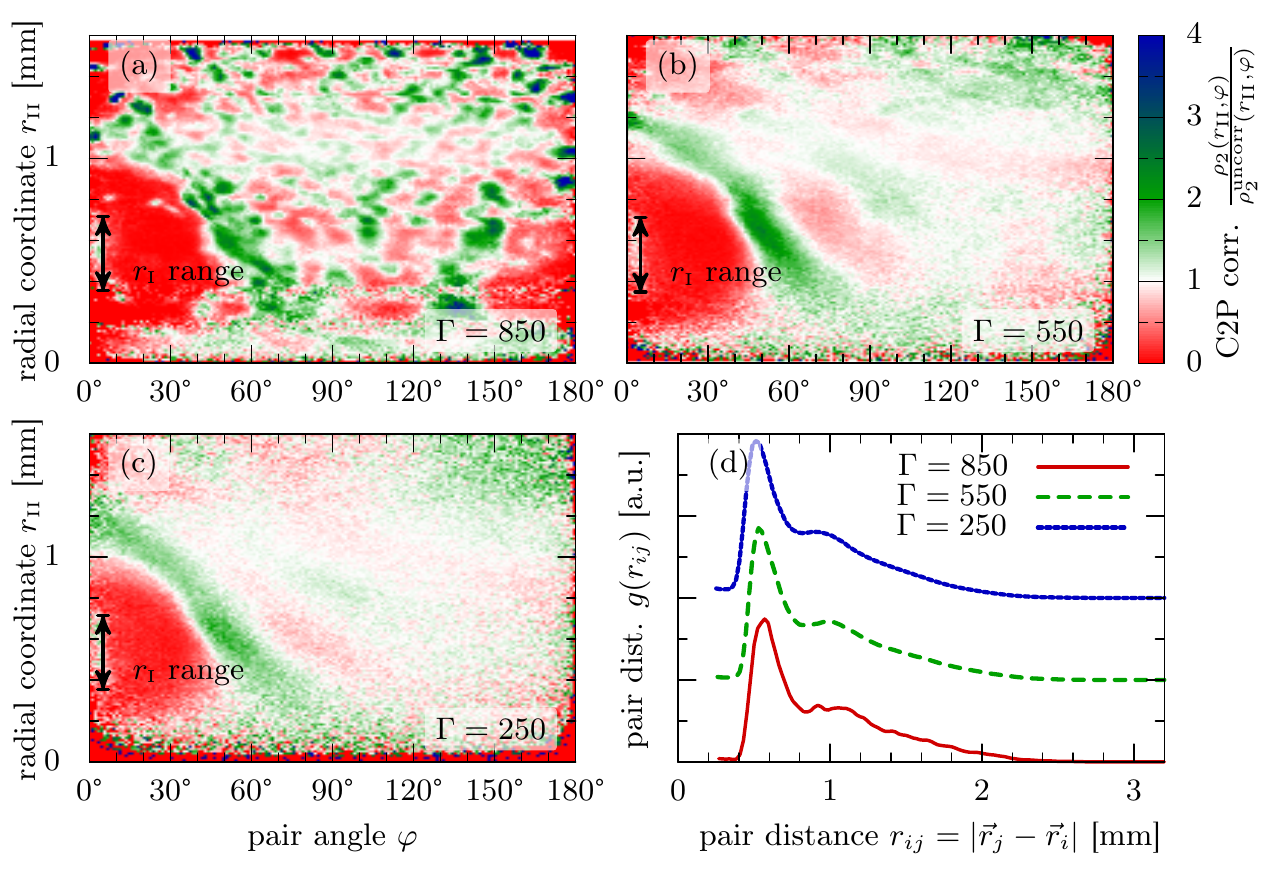}\\
  \caption{Center-two-particle correlation functions for the $N=60$ cluster,
  a) without laser heating ($\Gamma=850$), b) for a laser heating power of 100\,mW ($\Gamma=550$)
  and c) 300\,mW ($\Gamma=250$). d) Pair distribution function. The smoothening of the curve with
  decreasing $\Gamma$ indicates the loss of order but, in contrast to the C2P correlation function 
  in a)-c), the pair distribution function cannot distinguish intra-shell from inter-shell
  correlations. (The zero line is shifted for $\Gamma = 550$ and $\Gamma = 250$,
  for sake of clarity.) }\label{fig:TCF}
\end{figure}

Here, as an example, the heating and melting of a 3D cluster with $N=60$ particles is
demonstrated\footnote{This cluster is formed from 4.86\,$\mu$m particles at a gas pressure of 6.4\,Pa
and at a rf power of 1.3\,W.} using the pair of opposing heating laser beams. The Yukawa ball is
spherical in shape and consists of two shells, see Fig.~\ref{fig:traj}. By increasing the laser
power, the amount of heating is increased and melting is achieved. For low laser heating power
the cluster remains in a solid-like arrangement, as seen from the particle trajectories. Stronger
random particle motion is excited at higher laser power where then frequent intra-shell and
inter-shell particle exchanges are seen. Hence, the cluster is apparently driven into the liquid
regime. The change of structural and thermodynamic properties that is induced by the lasers is
illustrated in Fig.~\ref{fig:TCF}(d). There we show the pair distribution function $g(r_{ij})$ for three
different heating powers. The differences in $g(r_{ij})$ between the three heating powers are only
small. All three curves show a pronounced first-neighbor maximum at $r_{ij}\approx 0.6\,{\rm mm}$,
a shoulder at $r_{ij}\approx 1.2\,{\rm mm}$ corresponding to second neighbors and a decay to zero.
While the curves for $\Gamma=250$ and $\Gamma=550$ appear smooth, a substructure is visible at high
coupling, $\Gamma=850$. 
The reason for the weak sensitivity is that $g(r_{ij})$ does not distinguish between intra-shell neighbors and neighbors on different shells. This leads to a smeared out structure of $g(r_{ij})$.

The loss of order is reflected more clearly in the C2P correlation function.
As for the 2D case, it allows to quantitatively 
assess the melting transition(s) for these 3D clusters. This is shown in Fig.~\ref{fig:TCF} for the same cluster
($N=60$) with or without laser heating. As for the 2D case, one sees pronounced peaks at distinct 
angles $\gangle$ and radii $\rtwo$ at low heating powers and
subsequent loss of correlations for increased heating (reduced $\Gamma$). Also, in these 3D
systems, laser heating provides a near-equilibrium heating scenario. A two-step melting
(orientational melting before radial melting) which is expected for finite clusters
\cite{Bedanov94,Schweigert95c,Apolinario07,Apolinario08} is experimentally identified also for
3D clusters \cite{Schella11}.

\begin{figure}
  \includegraphics[width=87mm]{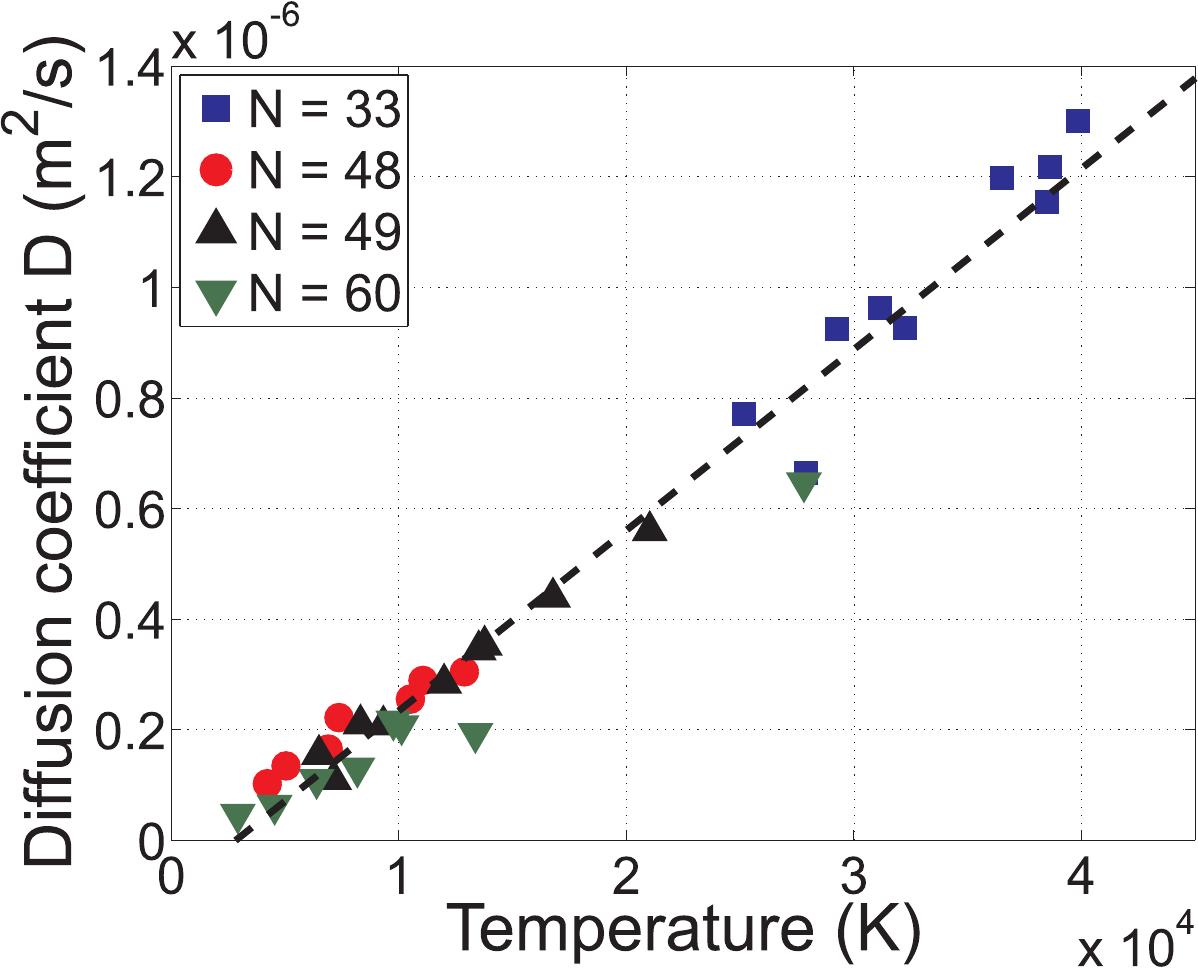}\\
  \caption{Diffusion coefficient $D(T)$ as a function of temperature for Yukawa balls of different
  size determined from an INM analysis. The dashed line corresponds to a linear fit to the diffusion
  coefficients for all clusters. (From Ref. \cite{Schella13}.)}\label{fig:diff}
\end{figure}

Finally, as another example of a liquid state property, the diffusion constant $D$, Eq.~(\ref{eq:diffusion-omega-integral}), has been derived from the experimental 3D particle trajectories of the $N=60$ cluster, based on the analysis of unstable instantaneous normal modes \cite{Melzer12,Melzer13,Schella13}, see Sec.~\ref{ss:transport}.
The so obtained diffusion coefficient is shown in Fig.~\ref{fig:diff}. For comparison, also the
diffusion constants of clusters of different sizes are added~\cite{Schella13}. A linear relationship
between $D$ and temperature is found for all studied clusters where $D$
 reaches values of about $D = 1.3 \cdot 10^{-6}$\,m$^2$/s at the highest dust
temperatures, $T\approx 4\cdot10^4$\,K. The values for the diffusion coefficient are decisively
larger than in the 2D case~\cite{Melzer12,Melzer13}. A reason for this is the higher
dimensionality of the system that allows more paths to change configurations~\cite{diff_coeff_note}.

As discussed in Sec.~\ref{sec:2Dexp}, from the diffusion constant an approximate melting
temperature can be extrapolated which is then found to be $T_M\approx2010$\,K for $N=60$.
Interestingly, this is smaller than the kinetic temperature of this cluster even
in the absence of heating which is $T_{\rm kin} = 2930$\,K \cite{Schella13}.
This relatively high kinetic temperature, even in the unheated case, is explained by
additional heating processes by the wake-field instability caused by streaming ions
in the plasma sheath~\cite{schweigert_alignment_1996,Schweigert98,Melandso97}.

\section{Discussion and outlook}\label{s:discus}
\subsection{Discussion of the results}
In this review, we discussed the various opportunities provided by laser beams as manipulation
tools for dusty plasmas. We focused on the use of lasers as heating instruments for dust
particles. As discussed in the introduction, various further uses, like the excitation of shear
stress~\cite{nosenko_shear_2004,feng_observation_2012} or of the rotation of a cluster
shell~\cite{klindworth_laser-excited_2000}, are possible as well, but this goes beyond the scope
of this review. The availability of reliable heating tools is essential for the experimental
investigation of phase transitions and instabilities in dust crystals. The enhanced manipulation
setup for finite 2D dust clusters, presented in Sec.~\ref{sec:general_concept}, has proven to be
usable as a tunable thermostat in both experiment and LMD simulation. The experiments
confirmed that the used laser heating scheme provides a homogeneous power input over the
entire cluster and over all frequencies. As desirable for a true thermal heating, the isotropy and
the Maxwellian shape of the velocity distribution are preserved to very high accuracy. Various
laser scanning concepts have been studied in the simulations allowing to predict the optimal
parameters for the experiments.

The laser heating method was used to perform temperature scans of small 2D dust clusters with
different particle numbers. The instantaneous normal mode analysis allowed us to calculate the
diffusion constant and, by this, to determine an approximate melting temperature. This melting
temperature is found to be crucially dependent on the exact particle number, as a consequence
of different cluster symmetries. At the same time the INM analysis does not allow to
discriminate between intra-shell and inter-shell melting. As another more sensitive quantity we
studied the center-two-particle correlation function which displays both 
intra-shell and inter-shell correlation.

The heating method for 3D dust clusters has to work with two opposing laser beams, due to the required space for diagnostics and illumination of the cluster.
Hence, the heating quality with respect to the isotropy and the Maxwellian shape of the velocity distribution is not as perfect as for 2D clusters.
Nevertheless, as in 2D, the quality of this heating technique is sufficient to manipulate the dust temperature (and hence the coupling strength) in a controlled manner, without affecting the plasma parameters. 
Due to the spherical symmetry of 3D Yukawa balls, structural parameters which take into account this symmetry are required in order to investigate the structure and the melting behavior of these systems.
The most sensitive quantity turned out to be the center-two-particle correlation function which displays both intra-shell and inter-shell correlation and allows us to distinguish solid clusters with a highly ordered structure from molten clusters.
While important aspects of the melting behavior in 3D are now understood, accurate data for the phase diagram and its the particle number dependence are still missing.
The presented experimental tool--laser heating--, combined with the diagnostic based on the center-two-particle correlation function should allow to study these questions in detail in the near future.

\subsection{Extension to transport properties of inhomogeneous strongly coupled systems}
\begin{figure}
 \includegraphics{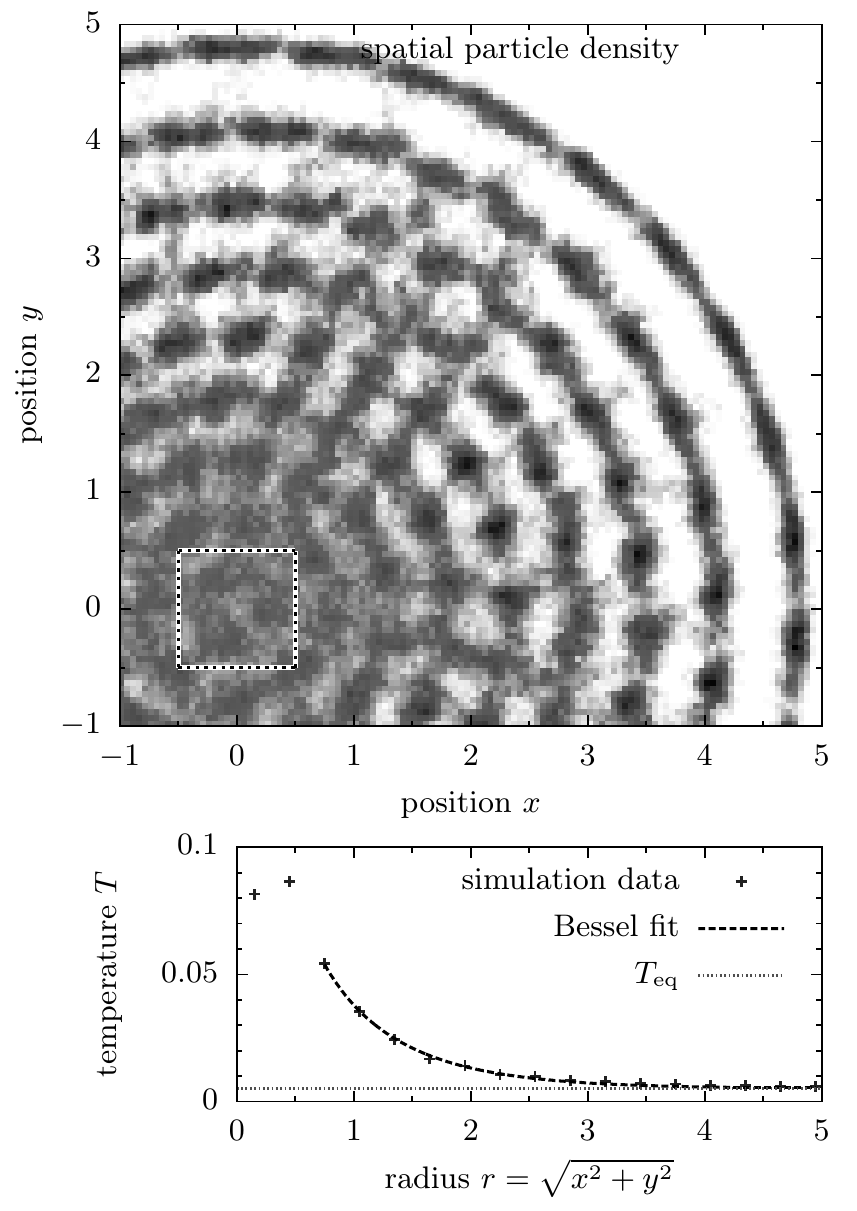}
 \caption{Top: Spatial density of a Yukawa cluster with $N=200$ particles. Four randomly moving 
 laser spots heat the inner square marked by the dashed rectangle.
 Bottom: Radial temperature profile and the fit by the analytical model presented 
 in~\cite{kudelis_heat_2013}. The temperature towards the border approaches the 
 equilibrium temperature $T_{\rm eq}$ of the unheated cluster.\label{fig:MDHeatTrans}}
\end{figure}

Both experiments and LMD simulations have shown that an appropriate heating scheme allows
for a homogeneous heating of the entire 2D
cluster~\cite{schablinski_laser_2012,thomsen_laser_2012}. However, this heating method can
also be used to heat only a selected spatial region. As long as the heated region is larger than
the laser spot size, local heating is simply realized by restricting the area which is scanned by
the laser spots. Kudelis \textit{et al.} suggested an experiment where only the central region of a
2D Yukawa cluster is heated by four randomly moving laser spots and performed LMD
simulations for this scenario~\cite{kudelis_heat_2013}. The thermal conductivity associated
with the radial temperature profile, see Fig.~\ref{fig:MDHeatTrans} (bottom), is found to be
constant over a wide range in coupling strengths, including the phase transition between solid
and liquid~\cite{kudelis_heat_2013}. This result is in good agreement with experimental data
for the heat transport in an extended 2D dust crystal by Nosenko \textit{et
al.}~\cite{nosenko_heat_2008}. The upper part of Fig.~\ref{fig:MDHeatTrans} shows the spatial
particle density of the Yukawa cluster that is heated in the inner region. While a pronounced
shell structure is found at the cold outer region, the center of the cluster is molten.

It remains an interesting task to verify these predictions in an experiment and, thus, employ
laser heating for the measurement of heat transport in strongly correlated finite dusty plasmas.
Another interesting task is to study the influence of an electric or magnetic field on the transport
coefficients.

\subsection{Control of time-dependent processes in complex plasmas}\label{ss:t-dep}
The applications discussed so far all considered stationary states. However, the laser heating
method also allows for the study of transient, time-dependent processes.
Compared to the relatively slow time scale of the heavy dust particles, the power 
input can be turned on and off practically immediately by switching on an off the external 
laser source.
A possible application is a temperature quench during which the system is 
abruptly cooled (or heated). This allows, for example, to study the time-dependent relaxation
from a fluid into a crystalline state.
Alternatively, quenching of a dust crystal was achieved by changing the properties of the ambient plasma,
e.g. via a sinusoidal modulation of the dc self-bias of the lower, powered electrode~\cite{Hartmann2010_PRL},
or by applying shock waves induced via a wire placed beneath the crystal~\cite{Knapek2007_PRL,Knapek2013_PRL}.
However, this alters the plasma environment and does not allow to vary only $\Gamma$ by a single parameter,
namely the dust kinetic temperature, like in the laser experiments presented in this review.

Feng \textit{et al.} \cite{Feng08} laser-heated, and subsequently rapidly
cooled, a 2D dust crystal that initially was arranged in a hexagonal lattice. During rapid heating,
the dust arrangement remained in a solid structure at temperatures above the melting point,
demonstrating solid superheating.
%
The relaxation of small, 2D dust clusters was investigated by means of lasers by 
Lisin \textit{et al}.~\cite{Lisin2013_NJP}. There and in Ref.~\cite{Knapek2007_PRL}, the cooling
rate for the crystallization of a 2D dust system was found to be close to the friction
coefficient, at moderate damping rates. A different behavior was found in laser-mediated
recrystallization experiments with small 3D dust clouds~\cite{Schella2014_PoP} where the cooling rate was found to be decisively 
lower than the friction coefficient for moderate damping (i.e. $\omega_0 \approx \gamma$),
confirming previous simulations~\cite{Kaehlert2010_PRL}. However, further investigations
are required to study the shell formation process in larger dust clouds with several
thousand particles and a multi-shell structure. A possible approach to circumvent the
obstacle of high particle numbers could be the use of tracer particles to artificially
reduce the particle density~\cite{Miksch07_PRE, Himpel2012_PoP}. 

Furthermore, since dust clusters are, at low neutral gas pressure, strongly influenced by wake effects due to fast 
streaming ions and thus exhibit attractive forces leading to particle chain formation~\cite{kroll2010,ludwig_njp}, it would be 
tempting to apply a laser-induced torque to the
dust system in order to study the energetic landscape by screwing the aligned dust ensemble.

A particularly interesting application would be a temperature quench in the presence of a strong magnetic field.
Ott \textit{et al.} found in MD simulations that a magnetic field may prevent crystallization. 
This is unexpected since, due to the Bohr von Leeuwen theorem, a magnetic field should not affect
the static properties of a classical system. The reason for the observed effect is that a strong magnetic
field may prevent conversion of potential energy into kinetic energy creating a bottleneck for a phase transition.
Above at critical magnetic field strength, the relaxation time $\tau_{\rm r}$ for the crystallization
increases exponentially~\cite{ott_magnetic_2013}.  It would be very interesting to verify these simulation results
in a laser heating experiment. So far magnetizing dust particles in an experiment has not been possible which is
due to the low specific charge $Q_{\rm d}/m$ of the dust grains.
An alternative way to effectively ``magnetize'' the dust component is to put the cluster into
rotation via a rotation of the neutral gas \cite{carstensen_pop12}. The Coriolis force then has the
same functional form, $\vec{F}_{\rm C} \propto \vec{v} \times \vec{\omega}$, as the Lorentz force
and acts as a ``pseudo-magnetic'' field (Larmor's theorem). This
idea has in fact been realized in dusty plasma experiments. It was shown that important
properties such as collective modes of magnetized strongly correlated plasmas can be
accurately reproduced~\cite{kaehlert_prl12,bonitz_magnetized_2013,hartmann_prl13}. In
combination with the laser heating (or cooling) method presented in this paper, this technique should allow to 
perform temperature quenches in experiments with finite dust clusters and to investigate the influence of the magnetization on the relaxation
time for crystallization.

This outlines just a few possible further directions of laser heating of strongly correlated finite dust clusters. Besides thermodynamic properties 
which were in the focus of this article, an experimental analysis of time dependent processes is now within reach.
The relevance of correlation effects in many fields of physics should make such studies interesting also beyond the field of dusty plasmas.

{\bf Acknowledgements.} We acknowledge financial support by the Deutsche Forschungsgemeinschaft via SFB-TR 24, grants A3, A7 and A9 and a grant for CPU time at the HLRN (grant SHP006), the HEPP and M. Mulsow for the help with data processing. 
%

\bibliographystyle{iopart-num}
\bibliography{2013_LaserReview,melzer}

\begin{thebibliography}{100}
\bibitem{tau-average}
The integral in the expression for the effective time scale can be solved using
  \textsc{Mathematica}.
\bibitem{tau_note}
$\tau_{\rm eff}$ is inversely proportional to the spot velocity, when both
  $v_{l,x}$ and $v_{l,y}$ intervals are scaled by the same factor. Increasing
  the laser scanned area has the side effect of decreasing $\tau_{\rm eff}$ via
  the faster spot velocity at constant scanner frequencies. Therefore, the
  temperature increaes scales as $\Delta T = T_{\rm eq}-T_{0} \propto X^{-3}$
  with the side length of the scanned area. At constant aspect ratio $\sx:\sy$,
  the passing time $\tau_{\rm eff}$ scales linearly with $\sx$.
\bibitem{diff_coeff_note}
From the relation of the diffusion coefficient to the mean squared
  displacements, Eq.~(\ref{eq:mean-square-displacement}), one would expect a
  dimensionality-related increase by a factor $3/2$ which, however, is not
  sufficient to explain the observed difference of the experimental values.
  This problem is still under investigation.
\end{thebibliography}


\end{document}